\documentclass[epj,nopacs]{svjour}
\usepackage{graphics}
\usepackage{float}
\usepackage{dcolumn}   
\usepackage{bm}        
\usepackage{amssymb}   
\usepackage{slashed}   
\usepackage{amsmath}   
\usepackage{verbatim}  
\usepackage{color} 
\usepackage{xcolor} 
\usepackage{subfig}

\usepackage{lipsum}
\usepackage{mathtools}
\usepackage{cuted}

\usepackage[bookmarks=true,colorlinks=true,linkcolor=blue,unicode=true]{hyperref}

\newcounter{JW}

\begin{document}

\title{ Particle Physics Violating Crypto-Nonlocal Realism}
\subtitle{European Physical Journal C volume {\bf 80}, Article number: 861 (2020).}
\author{Yu Shi\inst{1} \and Ji-Chong Yang\inst{1,2}}
\institute{Department of Physics, Fudan University, Shanghai, 200433, China \and Department of Physics, Liaoning Normal University, Dalian 116029, China}

\abstract{
It has been well established that quantum mechanics (QM) violates Bell inequalities (BI), which  are consequences of local realism (LR). Remarkably  QM also violates Leggett inequalities (LI), which  are consequences of  a  class of  nonlocal realism  called crypto-nonlocal realism (CNR).   Both LR and CNR assume that measurement outcomes are determined by  preexisting objective properties, as well as hidden variables (HV) not considered in QM.
We extend CNR and LI to include the case that the measurement settings are not externally fixed, but  determined by HV. We derive a new version of LI, which is then shown to be violated by entangled $B_d$ mesons,  if charge-conjugation-parity (CP) symmetry is indirectly violated, as indeed established. The experimental result is quantitatively estimated by using the indirect CP violation parameters,  and  the maximum  of a suitably defined relative violation is about $2.7\%$. Our work implies that   particle physics  violates CNR. Our LI can also be tested in other systems such as photon polarizations.
}


\maketitle

\section{\label{sec:level1}Introduction}

In 1935, Einstein, Podolsky and Rosen (EPR) questioned the completeness of QM, by applying  a criterion of LR to a pair of  particles in a quantum state which Schr\"{o}dinger subsequently referred to as entangled~\cite{EPR,sch}. Locality means that two events cannot  have any mutual physical influence if they are spacelike separated, that is, their spatial separation is larger than the distance the fastest physical signal, i.e. the light,  can travel within the time difference between the two events.   In 1964, Bell proposed the first BI satisfied by any local realistic theory while violated by QM~\cite{BI}. A  more experimentally suitable version of BI, called  Clauser-Horne-Shimony-Holt inequality~\cite{CHSHI}, was  demonstrated to be violated in many experiments, including   the ones  closing the locality loophole~\cite{BIAspect,z1}, the detection loophole~\cite{d1,d2},  and both~\cite{BIfree1,free2,free3}. To close yet another loophole called  measuring setting or freedom of choice loophole, observations of Milky Way stars~\cite{as,as2} and human choices~\cite{bigbell} have been employed. Great progress has been made in making use of quantum entanglement in quantum information science.

With  the conflict between LR and QM  well established, it is important to identify which aspects of LR are the sources of the conflict.  For this purpose, Leggett in 2003 proposed the LI, which is satisfied by CNR  and is violated by QM~\cite{LI}. This means that even nonlocal realism, at least a subset,  cannot avoid the conflict with QM, so the source of conflict seems to be more likely realism.  In 2007, a version of LI was  experimentally demonstrated to be  violated by using polarizations of entangled photons generated in spontaneous parametric down conversion, first under an additional assumption of rotational  invariance~\cite{LIExpNature}, then  without this assumption~\cite{Paterek,Branciard}. LI violation was also demonstrated by using polarizations of photons from  fibre-based source~\cite{Eisaman}, as well as  the orbital angular momenta of photons~\cite{Romero}. Similar phenomena were observed in different degrees of freedom of single particles~\cite{Hasegawa,Cardano}. Various extended discussions have also been made~\cite{comment,Colbeck,Branciard2,Oh}.

It is highly interesting to extend the investigations on BI and LI to particle physics, of which standard model (SM) is based on quantum field theory combining QM with special relativity, emphasizing causality  and using  local gauge principle to describe fundamental interactions. Massive and possibly unstable particles governed by strong and weak interactions and flying in relativistic velocities represent a new class beyond both photons and nonrelativistic particles governed merely by electromagnetism, and can still easily achieve spacelike separation. Besides, one might also wonder  whether high energy particles,  as excitations of quantum fields,  may display nonlocal effects.

In particle physics, entanglement, more often called EPR correlation,  has been noted  in pseudoscalar meson  pairs since 1960s~\cite{1960s}. Various discussions were made on its   rigorous verification~\cite{lot}, which was experimentally done  in  $K^0\bar{K}^0$ pairs produced in proton-antiproton annihilation~\cite{cern},  in  $\phi$ resonance~\cite{kloe1,kloe2}, as well as in   $B^0_d\bar{B}^0_d$ pairs produced in $\Upsilon (4S)$ resonance~\cite{go1,go2,bevan}. Entanglement is routinely used to tag mesons  by  identifying their entangled partners~\cite{kloe2,bevan,DT,Betad,Ablikim,detectionEfficience}. Moreover, entangled meson pairs are used in measuring various  parameters~\cite{Ablikim,detectionEfficience,parameter}, and studying  violations of
discrete symmetries~\cite{bernaB,Bd,JDROtherReview}, including CP~\cite{lipkin89,dunietz,ImprovedCPV,bernaB,JDRxzz}, time reversal (T)~\cite{TViolation,TViolationD}, and CPT~\cite{bernabeucpt}. A possible scheme of teleporting mesons was also proposed~\cite{telep}.  There exists similar entanglement in hyperon pairs generated from electron-positron annihilation, which was  used recently to measure the phase between the amplitudes of the  decays to different helicity states~\cite{bes3}.

Many proposals had been made on BI test in entangled mesons~\cite{mesonbi,mesonbi2}, and in the analogous spin-entangled particles~\cite{toernqvist,privitera}. There had been  an early  experiment using entangled protons to test BI under a few additional assumptions~\cite{lamehi}. There was an experiment using entangled $B^0_d\bar{B}^0_d$  pairs to test BI, in which meson decay  acts as effective measurement settings~\cite{go1}. However, it was not regarded as a genuine Bell test,  because of the lack of active measurement~\cite{toernqvist,problemOfMeson,bevan}. Basically this is a manifestation of the  loophole of measurement settings, for the following reason.  One can envisage a local HV (LHV) theory in which HVs in the source of the  particle pairs determine the  times, modes and even products of the decays, and the information is carried  by the particles, consequently  the two particles are secretely correlated no matter how far away they are separated, rendering the violation of BI. Other approaches to BI using entangled high energy particles are difficult to realize, as the alternative bases of measurement are physically  limited.

In this paper, we extend  CNR and LI to include the case that the measurement settings are not externally fixed, but determined by HV, therefore the above situation jeopardizing  BI test in entangled mesons is allowed in CNR, and we propose LI test using entangled neutral $B_d$  mesons. From QM calculation of single particle decays, we identify the time-dependent  effective measuring directions due to the decays, as  counterparts of the directions of the polarizers measuring the photon polarizations.  For different decay times, they all  lie on a plane and a cone, respectively.  For such effective measuring directions, whether it is externally fixed or emerge from averages of measurement outcomes over HV,  we derive a new version of LI, which is violated by QM and entangled $B_d$ mesons. We calculate the measurable quantities  characterizing the relative magnitude of the LI violation, and find their maxima to be  about $2.7\%$. It turns out  that the LI can only be violated when CP symmetry is violated indirectly, i.e. in the mass matrix. Our work establish the true randomness of particle decay, including its time, mode, and product. On the other hand, our new LI can also be tested in other systems such as photon polarizations.

\section{Pseudoscalar Neutral Mesons \label{rev} }

In QM, a neutral pseudoscalar meson $M$ can be regarded as living in a two-dimensional Hilbert space, with basis states $|M^0\rangle$ and $|\bar{M}^0\rangle$, which are flavor eigenstates and  mutual CP conjugates, i.e. $CP|M^0\rangle = |\bar{M}^0\rangle$, $CP |\bar{M}^0\rangle =  |M^0\rangle$.  In this basis, the mass matrix is
 \begin{equation}
H\equiv {\bf M}-\frac{i}{2}{\bf \Gamma }=\left(
\begin{array}{cc}
H_{00} & H_{0\bar{0}} \\
H_{\bar{0}0} & H_{\bar{0}\bar{0}}
\end{array}\right),
\end{equation}
where   $H_{00}\equiv \langle M^0|H|M^0\rangle$, $H_{0\bar{0}}\equiv \langle M^0|H|\bar{M}^0\rangle$, and so on. The eigenstates of $H$ are
\begin{equation}
\begin{split}
&|M_1\rangle=\frac{1}{\sqrt{|p|^2+|q|^2}}[p|M^0\rangle + q|\bar{M}^0\rangle],\\
&|M_2\rangle=\frac{1}{\sqrt{|p|^2+|q|^2}}[p|M^0\rangle -  q|\bar{M}^0\rangle],
\end{split}
\end{equation}
with $p/q \equiv \sqrt{H_{\bar{0}0}/H_{0\bar{0}} }$.  The corresponding eigenvalues are
\begin{equation}
\begin{split}
\lambda _{1}&= m_{1}-\frac{i}{2}\Gamma _{1}=H_{00} +\sqrt{ H_{0\bar{0}}H_{\bar{0}0} },\\
 \lambda _{2}&= m_{2}-\frac{i}{2}\Gamma _{2}
 =H_{00} - \sqrt{ H_{0\bar{0}}H_{\bar{0}0} }.
\end{split}
\end{equation}

$H$ governs the evolution of the meson state
\begin{equation}|\psi(t)\rangle = a(t) |M^0\rangle+ b(t) |\bar{M}^0\rangle,
\end{equation}
with
\begin{equation}
\left(
\begin{array}{c}
a(t) \\
b(t) \\
\end{array}\right)
=U(t)
\left( \begin{array}{c}
 a(0)\\
 b(0) \\
\end{array}
\right),
\end{equation}
where
\begin{equation}
U(t) = \exp(-iHt)=  g_+(t)+g_-(t)
\left( \begin{array}{cc}
0&p/q \\
q/p&0 \\
\end{array}
\right),
\end{equation}
with  $g_{\pm}(t) \equiv \frac{e^{-i\lambda _2 t} \pm e^{-i\lambda _1  t}}{2}$. This leads  to the  mixing phenomena. Especially,  $M^0$ and $\bar{M}^0$ at $t=0$ evolve respectively to
\begin{equation}
\begin{split}
&|M^0(t)\rangle= g_+(t)|M^0\rangle
+\frac{q}{p}g_-(t)|\bar{M}^0\rangle,\\
&|\bar{M}^0(t)\rangle=
\frac{p}{q}g_-(t)|M^0\rangle+g_+(t)|\bar{M}^0\rangle.\\
\end{split}
\end{equation}

For a meson in an arbitrary state, its decay    to some final state $f$ indicates that there  has been a projection or filtering to some basis state $|\phi\rangle$, which decays to $f$,
$|\phi\rangle$ being~\cite{Bd}
\begin{equation}
|\phi\rangle = \frac{1}{\sqrt{|A_f|^2+ |\bar{A}_f|^2} } (\bar{A}_f^*|M^0\rangle+
A_f^*|\bar{M}^0\rangle), \label{phi}
\end{equation}
where $A_f=\langle f |W|M^0\rangle$ and $\bar{A}_f=\langle f |W|\bar{M}^0\rangle$ are,  respectively,  the  amplitudes of the decays from $M^0$ and  $\bar{M}^0$ to the final state $f$.  $ W = \mathcal{U}H_w$, where $H_w$ being the weak decay Hamiltonian while $\mathcal{U}$ being the strong evolution operator and reducing to the identity if  final state interactions are neglected.

A pair of neutral mesons can be produced as  $C=-1$ antisymmetric  entangled state
\begin{equation}
 |\Psi _- \rangle = \frac{1}{\sqrt{2}}\left(|M^0 \rangle |\bar{M}^0 \rangle - |\bar{M}^0 \rangle |M^0 \rangle \right),
 \end{equation}
where in each term, the first and second basis states are those of mesons $a$ and $b$ respectively.

Suppose this two-particle state evolves up to $t_a$, when meson $a$ decays to some final state $f_a$, indicating that there  is a projection or filtering of $a$ to some basis state $|\phi\rangle_a$, which decays to $f_a$. $|\phi\rangle_a$ is $|\phi\rangle$ as given in  (\ref{phi}) for meson a.  The meson $b$ continues to evolve till it decays to some final state $f_b$  at $t_b$,  indicating that there is a projection or filtering of $b$ to some basis  state  $|\phi\rangle_b$, which decays  to $f_b$.  $|\phi\rangle_b$ is $|\phi\rangle$ as given in  (\ref{phi}) for meson b. The time evolution of the entangled state up to the projections can be described as $
P_bU_b(t_b-t_a)P_aU_b(t_a)U_a(t_a)|\Psi _-\rangle
= P_bP_aU(t_b)U(t_a)|\Psi_-\rangle
= P_b P_a|\Psi_-(t_a,t_b) \rangle$,
where $P_a =|\phi\rangle_{aa} \langle \phi|$ and $P_b=|\phi\rangle_{bb} \langle \phi|$ are projection operators, and   the commutativity between operators on $a$ and those on $b$ have been used.  This justifies the  usual use of a state vector with two time variables
\begin{equation}
\begin{split}
&|\Psi _-(t_a,t_b)\rangle \equiv U(t_b)U(t_a)|\Psi_-\rangle \\
&= \frac{1}{\sqrt{2}}\left(|M^0(t_a)\rangle |\bar{M}^0(t_b)\rangle - |\bar{M}^0(t_a)\rangle |M^0(t_b)\rangle \right),
\end{split}
\label{eq.2.3}
\end{equation}
which means that the two mesons decay at $t_a$ and $t_b$, respectively.

Specifically, we use neutral $B_d$ mesons, because of the advantage that  $\Gamma_2 \approx \Gamma _1$, $q/p\approx e^{2i\beta}$, where $2\beta$ is a phase factor, $\beta$ is   given as  $\sin(2\beta)=
0.695$~\cite{PDG}.  Then  $M^0=B^0$, $\bar{M}^0=\bar{B}^0$,  $M_1=B_L$, $M_2=B_H$, and  $U(t)$ is simplified to
\begin{equation}
\begin{split}
&U(t)  =e^{-iMt-\frac{\Gamma}{2} t}\\
&\times \left( \cos\frac{x\Gamma t}{2}+i\sin \frac{x\Gamma t}{2} \left[ \cos(2\beta)\sigma ^x+\sin (2\beta)\sigma ^y\right]\right),
\end{split}
\label{bu}
\end{equation}
where $\sigma ^{i}$, $(i=x,y,z)$, are Pauli operators, $x \equiv (m_H-m_L)/\Gamma$, $M \equiv ( m_H+m_L)/2$ and $\Gamma \equiv (\Gamma_L+\Gamma_H)/2$, the subscripts following  those of $B_H$ and $B_L$.

In Bloch  representation, $|B^0\rangle$, like the horizontally polarized state  of a photon or the spin-up state of an electron, is represented as the vector $(0,0,1)$, while $|\bar{B}^0_d\rangle$, like the vertically   polarized state  of a photon or the spin-down state of an electron, is represented as the vector $(0,0,-1)$.  They can be chosen  as the  ``measuring directions'' or bases of measurement.

However, for a measurement following time evolution, it is more convenient to define an effective time-dependent basis or ``measuring direction''. A state  of a two-state system can be parameterized as
\begin{equation}
 |{\bf u} \rangle =e^{i\zeta} \left( \cos \frac{\theta_{\bf u}}{2}|0\rangle + e^{i\rho_{\bf u}} \sin \frac{\theta_{\bf u}}{2} |1\rangle\right),
\label{eq.2.1}
\end{equation}
where $|0\rangle$ and $|1\rangle$ represent the basis states. We consider its  time evolution that can be  parameterized   as
\begin{equation}
U(\theta _{\bf a},\rho _{\bf a})=\left(\cos\frac{\theta _{\bf a}}{2}-i\sin \frac{\theta _{\bf a}}{2}\left(\cos(\rho _{\bf a})\sigma ^x+\sin (\rho _{\bf a})\sigma ^y\right)\right),
\label{eq.2.2}
\end{equation}
of which (\ref{bu}) is an example, multiplied by  an additional decay factor  $e^{-iMt-\frac{\Gamma}{2} t}$.

Suppose that following the evolution $U(\theta _{\bf a},\rho _{\bf a})$, a signal   is recorded  as $\mathcal{A}=+1$ if $|0\rangle$ is detected, while $\mathcal{A}=-1$ if $|1\rangle$ is detected.  The QM expectation value of $\mathcal{A}$ is
\begin{equation}
\begin{split}
&\bar{\mathcal{A}}({\bf u})=\frac{|\langle  0|U|{\bf u}\rangle|^2-|\langle 1|U|{\bf u}\rangle|^2}{|\langle 0|U|{\bf u}\rangle|^2+|\langle 1|U|{\bf u}\rangle|^2}={\bf u}\cdot {\bf a},
\end{split}
\label{eq.3.qm}
\end{equation}
where
\begin{equation}{\bf u} =\left( \sin \theta_{\bf u}\cos \rho_{\bf u},  \sin \theta_{\bf u} \sin \rho_{\bf u}, \cos \theta_{\bf u}\right)\end{equation}
is the Bloch vector of  $|{\bf u} \rangle$, while
\begin{equation}{\bf a} = \left( -\sin \theta_{a}\sin \rho_{a},  \sin \theta_{a} \cos \rho_{a}, \cos \theta_{a}\right)\end{equation}
is the Bloch vector of  $U^\dagger(\theta _{\bf a},\rho _{\bf a})  |0\rangle$.
This can be easily understood by   regarding $\bar{\mathcal{A}}({\bf u})$   as expectation value of the signal obtained by measuring the initial state $|{\bf u}\rangle$ in the rotated basis $\{U^\dagger|0\rangle, U^\dagger|1\rangle\}$. The rotation $U^\dagger$ of the basis is   realized by evolution.

For a $B_d$ meson, the measurement in the flavor basis \{$|B^0\rangle$, $|\bar{B}^0\rangle$\}, corresponding to $\mathcal{A}_l=\pm1$, can be made through the semileptonic decay channel,  as the direct CP violation or wrong sign decay is negligible~\cite{PDG}. Thus $|\phi\rangle$ in (\ref{phi}) reduces to  $|B^0\rangle$ or $|\bar{B}^0\rangle$, and  one can define
\begin{equation}
\begin{split}
&\bar{\mathcal{A}}_l({\bf u})=\frac{|\langle B^0|U(t)|{\bf u} \rangle|^2-|\langle \bar{B}^0|U(t)|{\bf u}\rangle|^2}{|\langle B^0|U(t)|{\bf u}\rangle|^2+|\langle \bar{B}^0|U(t)|{\bf u}\rangle|^2}={\bf u} \cdot {\bf a}^l(t),
\end{split}
\label{eq.3.3l}
\end{equation}
where
\begin{equation*}
{\bf a}^l (t)=\left(\sin(2 \beta) \sin( x\Gamma t), -\cos(2 \beta) \sin(x\Gamma t), \cos(x\Gamma t)\right).
\end{equation*}

Likewise,   as the direct CP violation is negligible~\cite{PDG}, if the decay products are CP eigenstates $S_{\pm}$,  they signals the projection of  the  meson to CP basis states  $|B_{\pm}\rangle \equiv  \left(|B^0\rangle \pm |\bar{B}^0\rangle\right)/\sqrt{2}$. In this case, $|\phi\rangle$ in (\ref{phi}) reduces to  $|B_\pm\rangle$.

With $B_{\pm}$ corresponding to  $\mathcal{A}_s=\pm 1$, one can define
\begin{equation}
\bar{\mathcal{A}}_s({\bf u})=\frac{|\langle B_+|U(t)|{\bf u}\rangle|^2-|\langle B_-|U(t)|{\bf u}\rangle|^2}{|\langle B_+|U(t)|{\bf u}\rangle|^2+|\langle B_-|U(t)|{\bf u}\rangle|^2}={\bf u}\cdot {\bf a}^s(t),
\label{eq.3.3s}
\end{equation}
where
\begin{equation*}
\begin{split}
&{\bf a}^s (t) =\left(\sin ^2(2 \beta) \cos (x\Gamma t)+\cos ^2(2 \beta),\right.\\
&\left.\sin (4 \beta) \sin ^2\left(x\Gamma t/2\right),-\sin (2 \beta) \sin (x\Gamma t)\right).
\end{split}
\end{equation*}
Eqs.~(\ref{eq.3.3l}) and (\ref{eq.3.3s}) are of the same form as the standard QM result (\ref{eq.3.qm}) because the factor $e^{-\Gamma t}$ exists in all terms  in both denominator and the numerator, and thus cancels.

Note that $|\langle B^0|U(t)|{\bf u} \rangle|^2$, $|\langle \bar{B}^0|U(t)|{\bf u}\rangle|^2$ and $|\langle B_\pm|U(t)|{\bf u}\rangle|^2$ do not depend on the specific decay channels signalling the projection, hence  is not directly observed. In contrast CP asymmetries usually defined
depend on which channels are observed.

However, under the assumption of no wrong sign decays and no direct CP violation, $|\langle B^0|U(t)|{\bf u} \rangle|^2$, $|\langle \bar{B}^0|U(t)|{\bf u}\rangle|^2$ and $|\langle B_\pm|U(t)|{\bf u}\rangle|^2$ and thus the asymmetries we define  above are  related to the rates of decays  in specific channels. For  flavor eigenstates  $l^{\pm}$, as the final states of semileptonic decays,   $$\langle l^+|WU(t)|{\bf u} \rangle= \langle l^+|W |B^0\rangle  \langle B^0|U(t)|{\bf u} \rangle=A_{l^+}\langle B^0|U(t)|{\bf u} \rangle,$$ $$\langle  l^-|WU(t)|{\bf u} \rangle=\langle l^-|W | \bar{B}^0\rangle \langle \bar{B}^0|U(t)|{\bf u}\rangle=\bar{A}_{l^-}\langle \bar{B}^0|U(t)|{\bf u}\rangle.$$
For $|l^-\rangle=CP|l^+\rangle$, we have $A_{l^+}=\bar{A}_{l^-}=A_l$.
For CP eigenstates $S_\pm$ as the final states,  $$ \langle S_+|WU(t)|{\bf u}\rangle = \langle S_+|W|B_+\rangle \langle B_+|U(t)|{\bf u}\rangle,$$  $$ \langle S_-|WU(t)|{\bf u}\rangle = \langle S_-|W|B_-\rangle\langle B_-|U(t)|{\bf u}\rangle.$$
Note that here there is no special relation between  $l^+$ and $l^-$, or  between $S_+$ and $S_-$, as different decay channels may signal a same projection  in the meson Hilbert space. The examples of flavor eigenstates $l^{\pm}$  as the final states include $M^-e^+\nu$, $M^+ e^-\bar{\nu}$, $M^-\mu^+\nu$, $M^+ \mu^-\bar{\nu}$,  etc. The examples of CP eigenstates $S_\pm$ as the final states include $J/\psi K_S$, $J/\psi K_L$, $KK$, $KKK$,  $\pi\pi$, $\pi\pi\pi$, $DD$, etc.

Therefore, the asymmetries $\bar{A}_{l}$ and $\bar{A}_{s}$,  defined above in the meson Hilbert space, can be obtained  from experimentally measurable quantities,
\begin{equation*}
\begin{split}
&\bar{\mathcal{A}}_l({\bf u})=\frac{\frac{|\langle l^+|WU(t)|{\bf u} \rangle|^2}{|\langle l^+|W|B^0\rangle |^2}-\frac{|\langle  l^-|WU(t)|{\bf u}\rangle|^2}{|\langle l^-|W|\bar{B}^0|^2}}{\frac{|\langle l^+|WU(t)|{\bf u} \rangle|^2}{|\langle l^+|W|B^0\rangle |^2}+\frac{|\langle l^-|WU(t)|{\bf u}\rangle|^2}{|\langle l^-|W|\bar{B}^0|^2}},\\
&\bar{\mathcal{A}}_s({\bf u})=\frac{\frac{|\langle S_+|WU(t)|{\bf u}\rangle|^2}{|\langle S_+|W|B_+\rangle|^2}-\frac{|\langle S_-|WU(t)|{\bf u}\rangle|^2}{|\langle S_-|W|B_-\rangle|^2}}{\frac{|\langle S_+|WU(t)|{\bf u}\rangle|^2}{|\langle S_+|W|B_+\rangle|^2}+\frac{|\langle S_-|WU(t)|{\bf u}\rangle|^2}{|\langle S_-|W|B_-\rangle|^2}}.
\end{split}
\end{equation*}

\begin{figure}
\resizebox{1.0\hsize}{!}{\includegraphics*{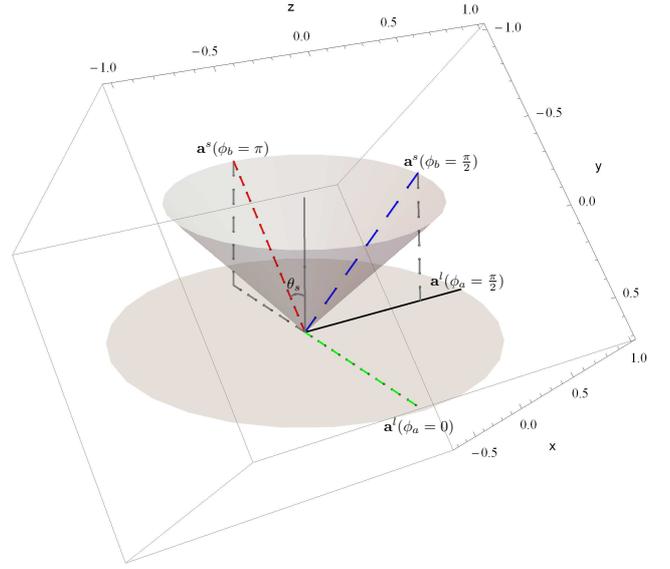}}
\caption{\label{fig:Polarizer}  The effective measuring  directions  ${\bf a}^l$  and ${\bf a}^s$. In a certain  coordinate system, ${\bf a}^l(\phi_l)$  is  on $xy$ plane,  ${\bf a}^s(\theta_s,\phi_s)$ is on a cone.  For $B_d$ mesons,  $\phi_l = x\Gamma t$, $(\phi_s, \theta_s)=(x\Gamma t+\pi/2,2\beta)$,  corresponding to flavor and CP measurements following evolution of time $t$, respectively.  For photon polarizations, ${\bf a}^l$ and  ${\bf a}^s$ are polarizer directions in  Bloch representation, and can be adjusted directly. }
\end{figure}

As shown in Fig.~\ref{fig:Polarizer}, with the time passing,  ${\bf a}^l(t)$ rotates on a plane, while ${\bf a}^s(t)$ rotates on a cone whose axis is perpendicular to ${\bf a}^l$ plane. For convenience, we adopt a new coordinate system in which ${\bf a}^l$ plane is the $xy$ plane, then
\begin{equation}
\begin{split}
&{\bf a}^l(\phi_l)=\left(\cos \phi_l,\sin \phi_l,0\right),\\
&{\bf a}^s(\theta _s,\phi_s)=\left(\sin\theta _s \cos \phi_s,\sin \theta _s\sin\phi_s,\cos \theta _s \right),
\end{split}
\label{eq.3.8}
\end{equation}
where $\phi_l= x\Gamma t$ and $\phi_s=x\Gamma t+\pi/2$
are the azimuthal  angles of  ${\bf a}^l$  and ${\bf a}^s$, respectively, $\theta_s=2\beta$   is the polar  angle of   ${\bf a}^s$, and it suffices to consider $0< \theta _s\leq \pi/2$.

${\bf a}^l(\phi_l)$ and ${\bf a}^s(\theta _s,\phi_s)$ are two effective measurement settings or  ``measuring directions''. For $B_d$ mesons, they are time-dependent.  The rotation of   basis or measuring direction  realized by evolution  explains the similarity between decay time and polarizer angle.  But ${\bf a}^l(\phi_l)$ and ${\bf a}^s(\theta _s,\phi_s)$  can also be used, say, for photon polarization, by directly adjusting   $ \phi_l $  and $ (\theta _s,\phi_s)$ in experiments.

\section{CNHV  theories \label{sec:level3} }

$|{\bf u}\rangle$ is an eigenstate of  the Pauli operator $\bm{\sigma}\cdot \mathbf{u}$   in the direction of $\mathbf{u}$. A particle in this state has a definite    $\mathbf{u}$.  For a single particle, QM result  can be reproduced by a  realistic or  HV theory, in which the  measurement outcomes are determined by preexisting  properties independent of the measurement, or ``elements of reality'' in the words of EPR.  Thus  $\mathbf{u}$ is  identified  as such an element of reality.

Consider  a HV theory. Suppose a particle with property ${\bf u}$ is measured along direction ${\bf a}$, then the dichotomic measurement  outcome $A=\pm 1$ is determined by the  hidden variables $\lambda$ in addition to the property  ${\bf u}$ and the local parameter ${\bf a}$.  This is also called a local realistic theory, in which the parameter ${\bf a}$ is local.  In  a nonlocal realistic theory,   $A$ also depends on a  non-local parameters, collectively denoted as  $\eta$.   In a crypto-nonlocal HV (CNHV)  theory, the individual properties  of each particle, after averaging  over distribution $\rho_{\mathbf{u} }(\lambda)$ of the hidden variables  $\lambda$,   become local, as indicated in  countless phenomena,
\begin{equation}
 \int d \lambda \rho_{\mathbf{u} }   (\lambda) A(\mathbf{u},\mathbf{a},\eta,\lambda)
 =\bar{A}(\mathbf{u}, \mathbf{a}). \label{ave}
\end{equation}
A concrete example of $\bar{A}(\mathbf{u}, \mathbf{a})$ is  the Malus' law~\cite{LIExpNature}
\begin{equation}
\bar{A}(\mathbf{u}, \mathbf{a})=\mathbf{u} \cdot\mathbf{a}, \label{avem}
\end{equation}
which is consistent with  QM results of photon polarizations. In the last section, we have just shown that it is also consistent with QM result of  the  meson  decay following its evolution.

For a pair of particles  from a common source,  with   respective properties $\mathbf{u}$ and $\mathbf{v}$, the measuring direction  of the other particle can serve  as the nonlocal parameter, and one can also assume nonlocal parameters $\eta_a$ and $\eta_a$, which are nonlocal with respect to a and b, respectively.   The  measurement outcomes  along  respective directions ${\bf a}$ and ${\bf b}$  are $  A(\mathbf{u},\mathbf{v},\mathbf{a},\mathbf{b},
\eta_a,\eta_b,\lambda)=\pm 1$ and $  B(\mathbf{v},\mathbf{u},\mathbf{b},\mathbf{a},
\eta_b,\eta_a,\lambda)=\pm 1$.  The local measurement of each particle cannot detect its correlation with the other particle, hence the nonlocal dependence disappears after averaging over the hidden variables,
\begin{equation}
\int d \lambda \rho_{\mathbf{u},\mathbf{v}}   (\lambda) A(\mathbf{u},\mathbf{v},\mathbf{a},\mathbf{b},
\eta_a,\eta_b,\lambda)
=\bar{A}(\mathbf{u}, \mathbf{a}), \label{abar}
\end{equation}
\begin{equation}
\int d \lambda \rho_{\mathbf{u},\mathbf{v}}   (\lambda) B(\mathbf{v},\mathbf{u},\mathbf{b},\mathbf{a},
\eta_b,\eta_a,\lambda)
=\bar{B}(\mathbf{v}, \mathbf{b}). \label{bbar}
\end{equation}
A general physical state is a statistical mixture of subensembles with definite    $ {\bf u}$ and ${\bf v}$.   Hence the final expectation values,  which is experimentally measured, are~\cite{LI,LIExpNature}
\begin{equation}
\begin{split}
&\langle A \rangle = \int d \mathbf{u} F(\mathbf{u}) \bar{A}(\mathbf{u}),\\
&\langle B \rangle = \int d \mathbf{v} F(\mathbf{v}) \bar{B}(\mathbf{u}),
\end{split}
\end{equation}
where $F(\mathbf{u})$ and $F(\mathbf{v})$ are probability distribution of polarizations $\mathbf{u}$ and $\mathbf{v}$, respectively. In case of correlated particles,  they are the reduced ones
\begin{equation}
\begin{split}
F(\mathbf{u})=&\int d\bf{v} F({\bf u},{\bf v}),   \\ F(\mathbf{v})=&\int d\bf{u} F({\bf u},{\bf v}).
\end{split}
\end{equation}

The two-body quantities may indicate correlations.  For definite  $ {\bf u}$ and ${\bf v}$,
\begin{equation}
\begin{split}
&\overline{AB}({\bf u},{\bf v},{\bf a},{\bf b})\\
&=\int d\lambda \rho _{{\bf u},{\bf v}}(\lambda) A({\bf u},{\bf v},{\bf a},{\bf b},\lambda) B({\bf v},{\bf u},{\bf b},{\bf a},\lambda).
\end{split}
\end{equation}
For a general state,
\begin{equation}
E({\bf a},{\bf b})=\int d{\bf u}d{\bf v}F({\bf u}, {\bf v})   \overline{AB}({\bf u},{\bf v},{\bf a},{\bf b}),
\end{equation}
which is  the main quantity to be investigated, as it may differ with the corresponding QM result  when entanglement is present, in which case a probability distribution over subensembles with definite polarizations leads to inequalities violated by the entangled state in QM.

Here we extend CNHV theories to include the case that ${\bf a}$ and ${\bf b}$ are not externally fixed. In each measurement, the  measurement settings $\tilde{\bf a}(\lambda)$ and  $\tilde{\bf b}(\lambda)$ are determined by HV $\lambda$, thus the measurement outcomes are  like $A(\mathbf{u}, \tilde{\bf a}(\lambda), \eta, \lambda)$.   Nevertheless, for  those measurements   with $\tilde{\bf a}(\lambda)= {\bf a} $, we can obtain the average of the outcomes. In the case of a single particle, the average is
\begin{equation}
\begin{split}
&\int d \lambda \rho'_{\mathbf{u},\mathbf{a} }(\lambda) A(\mathbf{u},\tilde{\bf a}(\lambda),\eta,\lambda)=\bar{A}(\mathbf{u}, \mathbf{a})=\mathbf{u} \cdot\mathbf{a},
\end{split}
\label{ave2}
\end{equation}
where $\rho'_{\mathbf{u},\mathbf{a} }(\lambda) \equiv \rho_{\mathbf{u} }(\lambda)\delta(\tilde{\bf a}(\lambda)-\mathbf{a})$ is a shorthand.

Likewise, for two correlated particles,   the outcomes of those measurements     with $\tilde{\bf a}(\lambda)= {\bf a} $ and   $\tilde{\bf b}(\lambda)= {\bf b}$ give rise to
\begin{equation}
\begin{split}
&\int d \lambda \rho_{\mathbf{u},\mathbf{v}} (\lambda)  \delta(\tilde{\bf a}(\lambda)-\mathbf{a})\delta(\tilde{\bf b}(\lambda)-\mathbf{b}) \\    & \times A(\mathbf{u},\mathbf{v},\tilde{\bf a}(\lambda),\tilde{\bf b}(\lambda),\eta_a,\eta_b,\lambda)\\
&=\bar{A}(\mathbf{u}, \mathbf{a})=\mathbf{u}\cdot \mathbf{a},
\end{split}
\label{abar1}
\end{equation}
\begin{equation}
\begin{split}
&\int d \lambda \rho_{\mathbf{u},\mathbf{v}} (\lambda)  \delta(\tilde{\bf a}(\lambda)-\mathbf{a})\delta(\tilde{\bf b}(\lambda)-\mathbf{b})  \\   & \times    B(\mathbf{v},\mathbf{u},\tilde{\bf b}(\lambda),\tilde{\bf a}(\lambda),\eta_b,\eta_a,\lambda)\\
&=\bar{B}(\mathbf{v}, \mathbf{b})=\mathbf{v}\cdot \mathbf{b}.
\end{split}
\label{bbar1}
\end{equation}
\begin{equation}
\begin{split}
&\overline{AB}({\bf u},{\bf v},{\bf a},{\bf b})=\int d\lambda \rho_{\mathbf{u},\mathbf{v}} (\lambda)  \delta(\tilde{\bf a}(\lambda)-\mathbf{a})\delta(\tilde{\bf b}(\lambda)-\mathbf{b})   \\
&\times A({\bf u},{\bf v},\tilde{\bf a}(\lambda),\tilde{\bf b}(\lambda),\eta_a,\eta_b,\lambda)
B({\bf v},{\bf u},\tilde{\bf b}(\lambda),\tilde{\bf a}(\lambda),\eta_b,\eta_a,\lambda).
\end{split}
\label{corab}
\end{equation}

Clearly the original formalism is  a special case of this extension, by externally fixing $\tilde{\bf a}(\lambda)$ to be always $\mathbf{a}$ and $\tilde{\bf b}(\lambda)$ to be always $\mathbf{b}$, independent of $\lambda$.

\section{\label{sec:level4}
LI for measuring directions on a plane and a cone}

We now  consider a pair of particles $a$ and $b$, with the measurement outcomes $A=\pm 1$ and $B=\pm 1$, respectively. The average  of those outcomes $A$  with a same measurement setting ${\bf a}$   satisfy the Malus' Law (\ref{abar1}).  The average  of those outcomes $B$  with a same measurement setting  ${\bf b}$ satisfy the Malus' Law (\ref{bbar1}).  The correlation function is defined  in the way of (\ref{corab}).   $\mathbf{a}$ and  $\mathbf{b}$ are each given as  ${\bf a}^l(\phi_i)$  or ${\bf a}^s(\theta_s,\phi_i)$, $(i=a,b)$, as in (\ref{eq.3.8}).

We first consider   correlation  functions of various combinations  of  ${\bf a}^l $  and ${\bf a}^s $ . Define $\hat{E}^\pm ({\bf a},{\bf b})\equiv E({\bf a},{\bf b} )+ E({\bf b},\pm{\bf b})$,  and rewrite $\hat{E}^{\pm}({\bf a}^s(\theta_s,\phi_a),{\bf a}^l(\phi_b))$ as $  \hat{E}^{\pm}_{sl}(\theta _s,\xi,\varphi)$, where $\xi \equiv (\phi _a+\phi _b)/2$,  $\varphi \equiv  \phi _a-\phi _b$. $\hat{E}^{\pm}_{ll}(\theta _s,\xi,\varphi)$ and $\hat{E}^{+}_{ss}(\theta _s,\xi,\varphi)$ are similarly  defined. Furthermore,  we consider the averages over $\xi$,
$\hat{E}^{-}_{sl} (\theta _s,\varphi)\equiv \int \frac{d\xi }{2\pi}  \hat{E}^{-}_{sl}  (\theta _s,\xi,\varphi)$ and so on.

In the Appendix, we prove the following  LI. The upper bound is given by
\begin{equation}
\begin{split}
& \hat{E}^{-}_{sl} (\theta _s,\varphi _1)+\frac{\pi \cos (\theta _{1}(\theta _s, \varphi _1)) L_1(\theta_s,\varphi _1) }{4\cos(\frac{\varphi _2}{2})} \hat{E}^{-}_{ll}(\theta _s,\varphi _2)  \\
&\leq 2  \left(1+\frac{\pi \cos (\theta _{1}(\theta _s, \varphi _1)) L_1(\theta_s,\varphi _1)  }{4\cos(\frac{\varphi _2}{2})}\right)\\
&-\cos (\theta _{1}(\theta _s, \varphi _1)) L_1(\theta_s,\varphi _1),
\end{split} \label{eq.3.25a}
\end{equation}
where
\begin{equation*}
\begin{split}
&L_1(\theta_s,\varphi)\equiv |{\bf a}^s+{\bf  a}^l|=\sqrt{2+2\cos(\varphi)\sin (\theta _s)},\\
&\theta _{1}(\theta_s,\varphi)=\cos ^{-1}\frac{\cos (\theta _s)}{\sqrt{2+2\cos (\varphi)\sin (\theta _s)}}.
\end{split}
\end{equation*}
With $0<\theta _s<\pi /2$, we have  $\sin(\theta _{1})> 0$, $\cos(\theta _{1})> 0$.

We find two lower bounds. The first is  given as
\begin{equation}
\begin{split}
&\hat{E}^+_{sl}(\theta _s,\varphi _1)+\frac{\pi \cos (\theta _{2}(\theta_s,\varphi _1))L_2(\theta_s,\varphi _1)}{4\left|\sin(\frac{\varphi _2}{2})\right|}\hat{E}^+_{ll}(\theta_s,\varphi _2) \\
&\geq -2\left(1+\frac{\pi \cos (\theta _{2}(\theta_s,\varphi _1)) L_2(\theta_s,\varphi _1)}{4\left|\sin(\frac{\varphi _2}{2})\right|}\right)\\
&+ \cos (\theta _{2}(\theta_s,\varphi _1)) L_2(\theta_s,\varphi _1).
\end{split}
\label{eq.3.29a}
\end{equation}
where
\begin{equation*}
\begin{split}
&L_2(\theta_s,\varphi)\equiv |{\bf a}^s-{\bf  a}^l|=\sqrt{2-2\cos(\varphi)\sin (\theta _s)},\\
&\theta _{2}(\theta_s,\varphi)=\cos ^{-1}\frac{\cos (\theta _s)}{\sqrt{2-2\cos (\varphi)\sin (\theta _s)}}.
\end{split}
\end{equation*}

The second lower bound is given as
\begin{equation}
\begin{split}
& \hat{E}^+_{sl}(\theta_s,\varphi _1)+\frac{\pi \cos (\theta _2(\theta_s,\varphi _1)) L_2(\theta_s,\varphi_1)}{4\sin(\theta _s)\left|\sin(\frac{\varphi _2}{2})\right|}\hat{E}^+_{ss}(\varphi _2) \\
&\geq -2\left(1+\frac{\pi \cos (\theta _2(\theta_s,\varphi _1)) L_2(\theta_s,\varphi_1)}{4\sin(\theta _s)\left|\sin(\frac{\varphi _2}{2})\right|}\right)\\
&+\cos (\theta _2)L_2(\theta_s,\varphi_1).
\end{split}
\label{eq.3.31a}
\end{equation}

Eqs.~(\ref{eq.3.25a}), (\ref{eq.3.29a}) and (\ref{eq.3.31a}) comprise  our LI.

The  correlation functions  averaged over $\xi$ are not directly observable, therefore rotational invariance   or  fair sampling of the averages needs to be assumed for measurements, in order that LI in terms of  these average correlation functions can be experimentally examined~\cite{Branciard,Paterek}. In the case of meson decays, the rotational invariance in Bloch representation  is actually time translational invariance.

To drop this additional assumption, we can redefine each  average in a discrete way,
\begin{equation}
\begin{split}
&\hat{E}^{\pm}_{sl}(\theta_s,\varphi)\equiv \frac{1}{N}\sum _{n=1}^N \hat{E}^{\pm}_{sl}(\theta_s,\xi_n =\frac{2n\pi}{N},\varphi),
\end{split}
\label{eq.3.43}
\end{equation}
and $\hat{E}^{\pm}_{ll}(\theta_s,\varphi)$ and  $\hat{E}^{+}_{ss}(\theta_s,\varphi)$ similarly.  As derived in the Appendix, for  these  discrete average correlation functions, our  LI can be obtained from Eqs.~(\ref{eq.3.25a}), (\ref{eq.3.29a}) and (\ref{eq.3.31a}) by simply replacing $\pi/4$ as $1/2u_N$, where  $u_N\equiv \frac{1}{N}\cot \left(\frac{\pi}{2N}\right)$. $N\geq 2$ is required. As  $N\to \infty$, $u_N\to 2/\pi$, then the discrete version  approaches the continuous version.

Our LI can be tested using various systems, in which measurement directions  ${\bf a}^l(\phi_l)$ and  ${\bf a}^s(\theta_s,\phi_s)$ can be directly adjusted.

For  meson decays,  $\theta_s=2\beta$ is fixed, while $\phi_l=\phi_l(t) = x\Gamma t$, $\phi_s=\phi _s(t) = x\Gamma t+\frac{\pi}{2}$ are  given by the decay time $t$. We mention that  for  the two particles $a$ and $b$  to be separated in spacelike distance, there is a constraint  on  the decay times $t_a$ and $t_b$. Suppose the particle pairs are generated from a particle at rest and each flies in velocity $v$ to opposite directions. Then spacelike separation means $(1+w)t_a > (1-w) t_b$, where $w=(v/c)/\sqrt{1-v^2/c^2}$.  Consequently there is a constraint on  possible values of $\xi$, but it does not affect the averages over $\xi$, which is an angle mathematically, hence its functions are periodic.

\section{\label{sec:level5}  Testing LI in entangled  \texorpdfstring{$B_d$}{B} mesons}

We now come back to the $C=-1$  $B^0\bar{B}^0$ entangled meson pairs, and we can  write the  correlation functions as
$$E({\bf a}^X(t_a),{\bf a}^Y(t_b)), (X,Y=l,s).$$
By definition,
\begin{equation}
\begin{split}
&E({\bf a}^X(t_a),{\bf a}^Y(t_b))= \\ & [(P(X+,Y+,t_a,t_b)+P(X-,Y-,t_a,t_b)\\ &-
P(X+,Y-,t_a,t_b)-P(X-,Y+,t_a,t_b)]/\\
&[P(X+,Y+,t_a,t_b)+P(X-,Y-,t_a,t_b)\\ &+
P(X+,Y-,t_a,t_b)+P(X-,Y+,t_a,t_b)],
\end{split}
\label{cor}
\end{equation}
where  for convenience, we have invented the shorthand
\begin{equation}
l+ \equiv B^0, \,\, l-\equiv \bar{B}^0, \, \, s\pm \equiv B_\pm,
\end{equation}
which should not be confused with $l^{\pm}$ and $S_{\pm}$, with $\pm$ as the superscript or subscript,  denoting  the final states of decays.

$P(X\pm,Y\pm,t_a,t_b)$ is the probability that the measurement results of a and b are $X\pm$ and $Y\pm$ at $t_a$ and $t_b$, respectively, indicated by the final states of their respective decays. The measurement or filtering or projection in the flavor basis $\{ B^0, \bar{B}^0\}$ is made through  a semileptonic decay to a flavor eigenstate $|l^\pm\rangle$.  The measurement or filtering  or projection  in CP basis $\{ B_+, B_-\}$ is made through a decay into CP eigenstate $|S_\pm\rangle$.  With direct CP violation negligible, we have $A_{l^-}=\bar{A}_{l^+}=0$,  $A_{S_{\pm}}=\pm \bar{A}_{S_{\pm}}$.  Only if $|l^-\rangle =CP |l^+\rangle$, we have $A_{l^+}=\bar{A}_{l^-}$. Note that the decay products of a and b may be different even though  their flavor  or CP eigenvalues are the same, and   may not be CP conjugates even though their flavor eigenvalues are opposites.

The basis of measurement, namely flavor or CP basis, is not actively  chosen  by experiments, but is signalled by the decay products.  Our extension   of CNHV and LI   addresses this issue, as the main achievement of this paper.

A key quantity is the joint decay rate for  particle   a  decaying  to    $f _a$ at $t_a$ while  particle  b  decaying to $f_b$ at $t_b$, $R(f _a,f _b, t_a, t_b) \propto \left|\langle f _a, f _b| W_a W_b | \Psi (t_a,t_b)\rangle \right|^2$. The following joint decay amplitudes  will be needed.

There are four amplitudes in the form of
\begin{equation}
\langle l^\pm_a, l^\pm_b| W_a W_b | \Psi (t_a,t_b)\rangle =
A_{l^\pm_a} A_{l^\pm_b} \langle  l \pm,  l\pm|   \Psi (t_a,t_b)\rangle, \label{ll}
\end{equation}
where,  $\pm$ in $l^\pm_a$ corresponds to the first  $ l \pm$ on RHS,  $\pm$ in   $l^\pm_b$ corresponds to the second $ l \pm$ on RHS.

There are four amplitudes of the form of
\begin{equation}
\begin{split}
&\langle S_\pm^a, S_\pm^b| W_a W_b | \Psi (t_a,t_b)\rangle =\\
&2 A_{S_\pm^a} A_{S_\pm^b} \langle s\pm, s\pm| \Psi (t_a,t_b)\rangle,
\end{split}
\label{ss}
\end{equation}
where we have used $\langle S_\pm|W|B_\pm\rangle =(A_{S_\pm}\pm \bar{A}_{S_\pm})/\sqrt{2}=\sqrt{2} A_{S_\pm }$.

There are four other amplitudes of the form of
\begin{equation}
\begin{split}
&\langle S_\pm^a, l^\pm_b| W_a W_b | \Psi (t_a,t_b)\rangle =\\
&\sqrt{2} A_{S_\pm^a} A_{l^\pm_b} \langle s\pm, l\pm| W_a W_b |  \Psi (t_a,t_b)\rangle.
\end{split}
\label{sl}
\end{equation}

The   experimentally measured quantity  is the  number of the joint events $N(f_a,f_b,t_a,t_b) \propto \epsilon_{f_a,f_b}R(f_a,f_b,t_a,t_b)$,
where $\epsilon_{f_a,f_b}$ is the detection efficiency for that channel~\cite{detectionEfficience}, $R(f_a,f_b,t_a,t_b)$ is proportional to  the modulo square of the joint decay amplitude, as given in   Eqs. (\ref{ll}-\ref{sl}).

Therefore the correlation function (\ref{cor}) can be obtained from event numbers as

\begin{equation}
\begin{split}
&E\left({\bf a}^l(t_a),{\bf a}^l(t_b)\right)\\
&=
\left(\frac{N(l_a^+,l_b^+,t_a,t_b)}{\epsilon_{l_a^+,l_b^+}|A_{l_a^+}|^2|A_{l_b^+}|^2}
+\frac{N(l_a^-,l_b^-,t_a,t_b)}{\epsilon _{l_a^-,l_b^-}|\bar{A}_{l_a^-}|^2|\bar{A}_{l_b^-}|^2}\right.\\
&\left.-\frac{N(l_a^+,l_b^-,t_a,t_b)}{\epsilon _{l_a^+,l_b^-}|A_{l_a^+}|^2|\bar{A}_{l_b^-}|^2}
-\frac{N(l_a^-,l_b^+,t_a,t_b)}{\epsilon _{l_a^-,l_b^+}|\bar{A}_{l_a^-}|^2|A_{l_b^+}|^2} \right)\\
&/\left(\frac{N(l_a^+,l_b^+,t_a,t_b)}{\epsilon _{l_a^+,l_b^+}|A_{l_a^+}|^2|A_{l_b^+}|^2}
+\frac{N(l_a^-,l_b^-,t_a,t_b)}{\epsilon _{l_a^-,l_b^-}|\bar{A}_{l_a^-}|^2|\bar{A}_{l_b^-}|^2}\right.\\
&\left.+\frac{N(l_a^+,l_b^-,t_a,t_b)}{\epsilon _{l_a^+,l_b^-}|A_{l_a^+}|^2|\bar{A}_{l_b^-}|^2}
+\frac{N(l_a^-,l_b^+,t_a,t_b)}{\epsilon _{l_a^-,l_b^+}|\bar{A}_{l_a^-}|^2|A_{l_b^+}|^2} \right),
\end{split}
\label{eq.4.8.1}
\end{equation}
\begin{equation}
\begin{split}
&E\left({\bf a}^s(t_a),{\bf a}^l(t_b)\right)\\
&=
\left(\frac{N(S_a^+,l_b^+,t_a,t_b)}{\epsilon _{S_a^+,l_b^+}|A_{S_a^+}|^2 |A_{l_b^+}|^2 }
+\frac{N(S_a^-,l_b^-,t_a,t_b)}{\epsilon _{S_a^-,l_b^-}|A_{S_a^-}|^2|\bar{A}_{l_b^-}|^2}\right.\\
&\left.-\frac{N(S_a^+,l_b^-,t_a,t_b)}{\epsilon _{S_a^+,l_b^-}|A_{S_a^+}|^2 |\bar{A}_{l_b^-}|^2 }
-\frac{N(S_a^-,l_b^+,t_a,t_b)}{\epsilon _{S_a^-,l_b^+}|A_{S_a^-}|^2  |A_{l_b^+}|^2 }\right)\\
&/\left(\frac{N(S_a^+,l_b^+,t_a,t_b)}{\epsilon _{S_a^+,l_b^+}|A_{S_a^+}|^2 |A_{l_b^+}|^2 }
+\frac{N(S_a^-,l_b^-,t_a,t_b)}{\epsilon _{S_a^-,l_b^-}|A_{S_a^-}|^2|\bar{A}_{l_b^-}|^2}\right.\\
&\left.+\frac{N(S_a^+,l_b^-,t_a,t_b)}{\epsilon _{S_a^+,l_b^-}|A_{S_a^+}|^2 |\bar{A}_{l_b^-}|^2 }
+\frac{N(S_a^-,l_b^+,t_a,t_b)}{\epsilon _{S_a^-,l_b^+}|A_{S_a^-}|^2  |A_{l_b^+}|^2 }  \right),
\end{split}
\label{eq.4.8.2}
\end{equation}
\begin{equation}
\begin{split}
&E\left({\bf a}^s(t_a),{\bf a}^s(t_b)\right)\\
&=
\left(
\frac{N(S_a^+,S_b^+,t_a,t_b)}{\epsilon _{S_a^+,S_b^+}|A_{S_a^+}|^2|A_{S_b^+}|^2}
+\frac{N(S_a^-,S_b^-,t_a,t_b)}{\epsilon _{S_a^-,S_b^-}|A_{S_a^-}|^2 |A_{S_b^-}|^2 }\right.\\
&\left.-\frac{N(S_a^+,S_b^-,t_a,t_b)}{\epsilon _{S_a^+,S_b^-}|A_{S_a^+}|^2|A_{S_b^-}|^2}
-\frac{N(S_a^-,S_b^+,t_a,t_b)}{\epsilon _{S_a^-,S_b^+}|A_{S_a^+}|^2|A_{S_b^+}|^2}\right)\\
&/\left(\frac{N(S_a^+,S_b^+,t_a,t_b)}{\epsilon _{S_a^+,S_b^+}|A_{S_a^+}|^2|A_{S_b^+}|^2}
+\frac{N(S_a^-,S_b^-,t_a,t_b)}{\epsilon _{S_a^-,S_b^-}|A_{S_a^-}|^2 |A_{S_b^-}|^2 }\right.\\
&\left.+\frac{N(S_a^+,S_b^-,t_a,t_b)}{\epsilon _{S_a^+,S_b^-}|A_{S_a^+}|^2|A_{S_b^-}|^2}
+\frac{N(S_a^-,S_b^+,t_a,t_b)}{\epsilon _{S_a^-,S_b^+}|A_{S_a^+}|^2|A_{S_b^+}|^2}\right),
\end{split}
\label{eq.4.8.3}
\end{equation}
where $\epsilon$'s are the detection efficiencies. In experiments, $|A_{l^{\pm}_{i}}|^2$ and  $|A_{S_{i}^{\pm}}|^2$, $(i=a,b)$,  can be absorbed to the redefinitions of detection efficiencies.

Furthermore, one obtains
\begin{equation}
\begin{split}
&\hat{E}^{ll\pm}(\phi_a,\phi_b) \equiv \\
&E\left({\bf a}^l(x\Gamma t_a),{\bf a}^l(x\Gamma t_b)\right)+E\left({\bf a}^l(x\Gamma t_b),\pm{\bf a}^l(x\Gamma t_b)\right),\\
&\hat{E}^{sl\pm}(\phi_a,\phi_b) \equiv \\
&E\left({\bf a}^s(x\Gamma t_a+\frac{\pi}{2}),{\bf a}^l(x\Gamma t_b)\right)+E\left({\bf a}^l(x\Gamma t_b),\pm{\bf a}^l(x\Gamma t_b)\right),\\
&\hat{E}^+_{ss}(\phi_a,\phi_b)  \equiv
E\left({\bf a}^s(x\Gamma t_a+\frac{\pi}{2}),{\bf a}^l(x\Gamma t_b)\right)\\&+E\left({\bf a}^s(x\Gamma t_b+\frac{\pi}{2}),{\bf a}^s(x\Gamma t_b+\frac{\pi}{2})\right),\\
\end{split}
\label{eq.4.8.new}
\end{equation}
from which the averages  $\hat{E}^{ll\pm}(\varphi)$, $\hat{E}^{sl\pm}(\varphi)$, $\hat{E}^{ss+}(\varphi)$  can be obtained.
Note that we did not define $\hat{E}^{ss-}$, which would not have physical meaning, as   $-{\bf a}^s$ is not on the cone, where all possible ${\bf a}^s$'s lie.
In SM, with $\Delta \Gamma = 0$, $R(f_b,f_b,t_a,t_b)$ can be obtained as 
\begin{equation}
\begin{split}
&R(f _a,f _b,t_a, t_b)=\frac{e^{-\Gamma (t_a+t_b)}}{4}\times \left\{(|\xi_-|^2+|\zeta_-|^2)\right.\\
&\left.-(|\xi_-|^2-|\zeta_-|^2)\cos (x\Gamma (t_a-t_b))\right.\\
&\left.-2Im(\zeta_-^*\xi_-)\sin (x \Gamma (t_a-t_b))\right\},
\end{split}
\label{eq.4.3}
\end{equation}
where $\xi_- \equiv -\left(\frac{p}{q}A_{f_a}A_{f_b}-
\frac{q}{p}\bar{A}_{f_a}
\bar{A}_{f_b}\right)$, $\zeta_- \equiv \left(A_{f_a}\bar{A}_{f_b}-
\bar{A}_{f_a}A_{f_b}\right)$.

In  experiments, it is more convenient to use the time-integrated joint decay rate
\begin{equation}
R(f _a,f _b,\Delta t)=\int _0^{\infty}dt_aR_-(f _a,f _b, t_a, t_a+\Delta t),
\label{eq.4.1}
\end{equation}
which is obtained as
\begin{equation}
\begin{split}
&R(f _a,f _b,\Delta t)= \frac{e^{-\Gamma |\Delta t|}}{8\Gamma}\times \left\{(|\xi_-|^2+|\zeta_-|^2)\right.\\
&\left. - (|\xi_-|^2-|\zeta_-|^2)\cos (x\Gamma \Delta t)\right.\\
&\left.+2Im(\zeta_-^*\xi_-)\sin (x\Gamma \Delta t)\right\},
\end{split}
\label{eq.4.3b}
\end{equation}

It is more rigorous to  test LI of  the discrete version of the average correlation functions, rather than that of the  continuous version. However, it is experimentally   much easier to measure   $N(f_a,f_b,\Delta t)\propto   \epsilon _{f_a,f_b}R(f_a,f_b,\Delta t)$  than   $N_{f_a,f_b}(t_a,t_b)$, consequently it is much easier to test LI in terms of  the continuous version of the average correlation functions.

From (\ref{eq.4.3}) and (\ref{eq.4.3b}), it can be seen that  in SM,  $R(f_a,f_b,t_b+\Delta t,t_b)=2\Gamma e^{-\Gamma (t_a+t_b)} e^{\Gamma |\Delta t|}    R(f_a,f_b, \Delta t)$. Consequently, $\hat{E}(\varphi)$, as an average of $E(\phi _a, \phi _b)$ over $\xi\equiv x\Gamma(t_a+t_b)/2+\pi/2$, can be directly related to $N(f_a,f_b,\Delta t)$ as
\begin{equation}
\begin{split}
&\hat{E}^{ll}(\varphi = x\Gamma \Delta t)
=\\
&\left(\frac{N(l_a^+,l_b^+,\Delta t)}{\epsilon_{l_a^+,l_b^+}|A_{l_a^+}|^2|A_{l_b^+}|^2}
+\frac{N(l_a^-,l_b^-,\Delta t)}{\epsilon _{l_a^-,l_b^-}|\bar{A}_{l_a^-}|^2|\bar{A}_{l_b^-}|^2}\right.\\
&\left.-\frac{N(l_a^+,l_b^-,\Delta t)}{\epsilon _{l_a^+,l_b^-}|A_{l_a^+}|^2|\bar{A}_{l_b^-}|^2}
-\frac{N(l_a^-,l_b^+,\Delta t)}{\epsilon _{l_a^-,l_b^+}|\bar{A}_{l_a^-}|^2|A_{l_b^+}|^2}\right)\\
&/\left(\frac{N(l_a^+,l_b^+,\Delta t)}{\epsilon _{l_a^+,l_b^+}|A_{l_a^+}|^2|A_{l_b^+}|^2}
+\frac{N(l_a^-,l_b^-,\Delta t)}{\epsilon _{l_a^-,l_b^-}|\bar{A}_{l_a^-}|^2|\bar{A}_{l_b^-}|^2}\right.\\
&\left.+\frac{N(l_a^+,l_b^-,\Delta t)}{\epsilon _{l_a^+,l_b^-}|A_{l_a^+}|^2|\bar{A}_{l_b^-}|^2}
+\frac{N(l_a^-,l_b^+,\Delta t)}{\epsilon _{l_a^-,l_b^+}|\bar{A}_{l_a^-}|^2|A_{l_b^+}|^2} \right),\\
\end{split}
\label{eq.4.9.1}
\end{equation}
\begin{equation}
\begin{split}
&\hat{E}^{sl}(\varphi = x\Gamma\Delta t+\frac{\pi}{2})=\\
&\left(\frac{N(S_a^+,l_b^+,\Delta t)}{\epsilon _{S_a^+,l_b^+}|A_{S_a^+}|^2 |A_{l_b^+}|^2 }
-\frac{N(S_a^-,l_b^-,\Delta t)}{\epsilon _{S_a^-,l_b^-}|A_{S_a^-}|^2|\bar{A}_{l_b^-}|^2}\right.\\
&\left.+\frac{N(S_a^+,l_b^-,\Delta t)}{\epsilon _{S_a^+,l_b^-}|A_{S_a^+}|^2 |\bar{A}_{l_b^-}|^2 }
-\frac{N(S_a^-,l_b^+,\Delta t)}{\epsilon _{S_a^-,l_b^+}|A_{S_a^-}|^2  |A_{l_b^+}|^2 }  \right)\\
&/\left(\frac{N(S_a^+,l_b^+,\Delta t)}{\epsilon _{S_a^+,l_b^+}|A_{S_a^+}|^2 |A_{l_b^+}|^2 }
+\frac{N(S_a^-,l_b^-,\Delta t)}{\epsilon _{S_a^-,l_b^-}|A_{S_a^-}|^2|\bar{A}_{l_b^-}|^2}\right.\\
&\left.+\frac{N(S_a^+,l_b^-,\Delta t)}{\epsilon _{S_a^+,l_b^-}|A_{S_a^+}|^2 |\bar{A}_{l_b^-}|^2 }
+\frac{N(S_a^-,l_b^+,\Delta t)}{\epsilon _{S_a^-,l_b^+}|A_{S_a^-}|^2  |A_{l_b^+}|^2 }  \right),\\
\end{split}
\label{eq.4.9.2}
\end{equation}
\begin{equation}
\begin{split}
&\hat{E}^{ss}(\varphi = x\Gamma\Delta t)
=\\
&\left(
 \frac{N(S_a^+,S_b^+,\Delta t)}{\epsilon _{S_a^+,S_b^+}|A_{S_a^+}|^2|A_{S_b^+}|^2}
+\frac{N(S_a^-,S_b^-,\Delta t)}{\epsilon _{S_a^-,S_b^-}|A_{S_a^-}|^2 |A_{S_b^-}|^2 }\right.\\
&\left.-\frac{N(S_a^+,S_b^-,\Delta t)}{\epsilon _{S_a^+,S_b^-}|A_{S_a^+}|^2|A_{S_b^-}|^2}
-\frac{N(S_a^-,S_b^+,\Delta t)}{\epsilon _{S_a^-,S_b^+}|A_{S_a^+}|^2|A_{S_b^+}|^2}\right)\\
&/\left(\frac{N(S_a^+,S_b^+,\Delta t)}{\epsilon _{S_a^+,S_b^+}|A_{S_a^+}|^2|A_{S_b^+}|^2}
+\frac{N(S_a^-,S_b^-,\Delta t)}{\epsilon _{S_a^-,S_b^-}|A_{S_a^-}|^2 |A_{S_b^-}|^2 }\right.\\
&\left.+\frac{N(S_a^+,S_b^-,\Delta t)}{\epsilon _{S_a^+,S_b^-}|A_{S_a^+}|^2|A_{S_b^-}|^2}
+\frac{N(S_a^-,S_b^+,\Delta t)}{\epsilon _{S_a^-,S_b^+}|A_{S_a^+}|^2|A_{S_b^+}|^2}\right).\\
\end{split}
\label{eq.4.9.3}
\end{equation}

Moreover,  the integration over $\xi$ of $\hat{E}^{\pm}({\bf a},{\bf b})=E({\bf a},{\bf b})+E({\bf b},\pm {\bf b})$  can be performed independently for the two  terms on RHS, consequently,
\begin{equation}
\begin{split}
&\hat{E}^+_{ll}(\varphi )=\hat{E}^{ll}(\varphi)+\hat{E}^{ll}(0),\;\;\hat{E}^{-}_{ll}(\varphi )=\hat{E}^{ll}(\varphi)+\hat{E}^{ll}(\pi)\\
&\hat{E}^+_{sl}(\varphi )=\hat{E}^{sl}(\varphi)+\hat{E}^{ll}(0),\;\;
\hat{E}^{-}_{sl}(\varphi )=\hat{E}^{sl}(\varphi)+\bar{E}^{ll}(\pi)\\
&\hat{E}^+_{ss}(\varphi )=\hat{E}^{ss}(\varphi)+\hat{E}^{ss}(0).\\
\end{split}
\label{eq.4.10.new}
\end{equation}

Note that Eqs.~(\ref{eq.4.8.1}-\ref{eq.4.8.3}) and (\ref{eq.4.9.1}-\ref{eq.4.9.3}) are mainly for the use in analyzing  experimental data. QM result  can be obtained simply from $\langle {\bf a}^X,{\bf a}^Y|\Psi_-\rangle=- {\bf a}^X\cdot{\bf a}^Y$, therefore
\begin{equation}
\begin{split}
& \hat{E}^{sl} (\varphi )=-\sin (2\beta)\cos (\varphi ),\;\; \hat{E}^{ll} (\varphi )=-\cos (\varphi ),\\
& \hat{E}^{ss }(\varphi )=-\cos^2(2\beta)-\sin^2(2\beta)\cos (\varphi ). \\
\end{split}
\label{eq.4.10}
\end{equation}
It is  interesting to test our LI using   various systems, in which $\varphi_a$ and  $\varphi_b$ are directly adjusted.

For $B_d$ mesons,  QM result (\ref{eq.4.10})  can also be obtained  from the definition (\ref{cor}) of correlation functions,   with \\ $P(X\pm,Y\pm,t_a,t_b)$ $\propto$ \\$\left|\langle X\pm,Y\pm| W_a W_b | \Psi (t_a,t_b)\rangle \right|^2$   obtained by substituting LHS  of $R(f _a,f _b, t_a, t_b) \propto$ $\left|\langle f _a, f _b| W_a W_b | \Psi (t_a,t_b)\rangle \right|^2$ with the result (\ref{eq.4.3}) of $R(f _a,f _b, t_a, t_b)$, and  RHS  with the joint amplitudes (\ref{ll}-\ref{sl}),  and then having
$A_f$ and  $\bar{A}_f$ cancelled.   For $B_d$ mesons,  the values of $\beta$ and $x$ are given by  $\sin (2\beta )=0.695$, $x=0.769$~\cite{PDG}.

The upper bound of a
our LI violation can be quantified as
\begin{equation}
g^u(\varphi_1,\varphi _2) \equiv \frac{h^u_{L}(\varphi_1,\varphi _2)-h^u_{R}(\varphi_1,\varphi _2)}{\left|h^u_{L}(\varphi_1,\varphi _2)\right|},
\end{equation}
where $h^u_{R}$ and $h^u_{L}(\varphi_1,\varphi _2)$ are RHS and LHS of (\ref{eq.3.25a}), respectively. The first lower bound can be quantified as
\begin{equation}
g^{d1}(\varphi_1,\varphi _2)=\frac{h^{d1}_{R}(\varphi_1,\varphi _2)-h^{d1}_{L}(\varphi_1,\varphi _2)}{\left|h^{d1}_{L}(\varphi_1,\varphi _2)\right|},
\end{equation}
where $h^{d1}_R$ and $h^{d1}_L$ are RHS and LHS of (\ref{eq.3.29a}), respectively.   The second  lower bound can be quantified as
\begin{equation}
g^{d2}(\varphi_1,\varphi _2)=\frac{h^{d2}_{R}(\varphi_1,\varphi _2)-h^{d2}_{L}(\varphi_1,\varphi _2)}{\left|h^{d2}_{L}(\varphi_1,\varphi _2)\right|},
\end{equation}
where $h^{d2}_R$ and $g^{h2}_L$ are RHS and LHS of  (\ref{eq.3.31a}), respectively. Each of these three quantities  larger than $0$ represents the violation of the corresponding bound of LI.

From
$\partial _{\varphi _1}g^u(\varphi _1,\varphi _2)= \partial _{\varphi _1}g^{d1}(\varphi _1,\varphi _2)= \partial _{\varphi _1}g^{d2}(\varphi _1,\varphi _2)=0$, it is determined that the maximum of   $g^{u}$ is on  $\varphi_1=\pi$, while the maxima of $g^{d1}$ and  $g^{d1}$ are both on $\varphi _1=0$.  $L_1(\varphi _1=\pi)=L_2(\varphi _1=0)\approx 0.781$,  $\theta _{1}(\varphi _1=\pi)=\theta _{ 2}( \varphi _1=0)\approx 0.401$.

Furthermore, solving $\partial _{\varphi _2}g^u(\pi,\varphi _2)=0$ numerically, we find that $g^u(\pi, \varphi _2)$ reaches its maximum at $\varphi _2\approx \pm 2.81$.  We also find numerically that  $g^u(\pi,\varphi _2)>0$ when    $2.39<|\varphi _2|<\pi$.
Similarly, solving $\partial _{\varphi _2}g^{d1}(0,\varphi _2)=0$ numerically, we find that $g^{d1}(0,\varphi _2)$ reaches the maximum at $\varphi _2\approx \pm 0.336$. We also find numerically  that $g^{d1}(0,\varphi _2)>0$ when    $ 0<|\varphi _2|<0.75$.

\begin{figure*}
\centering{
\subfloat[$g^u(\varphi _1, \varphi _2)$]{\resizebox{0.33\textwidth}{!}{\includegraphics{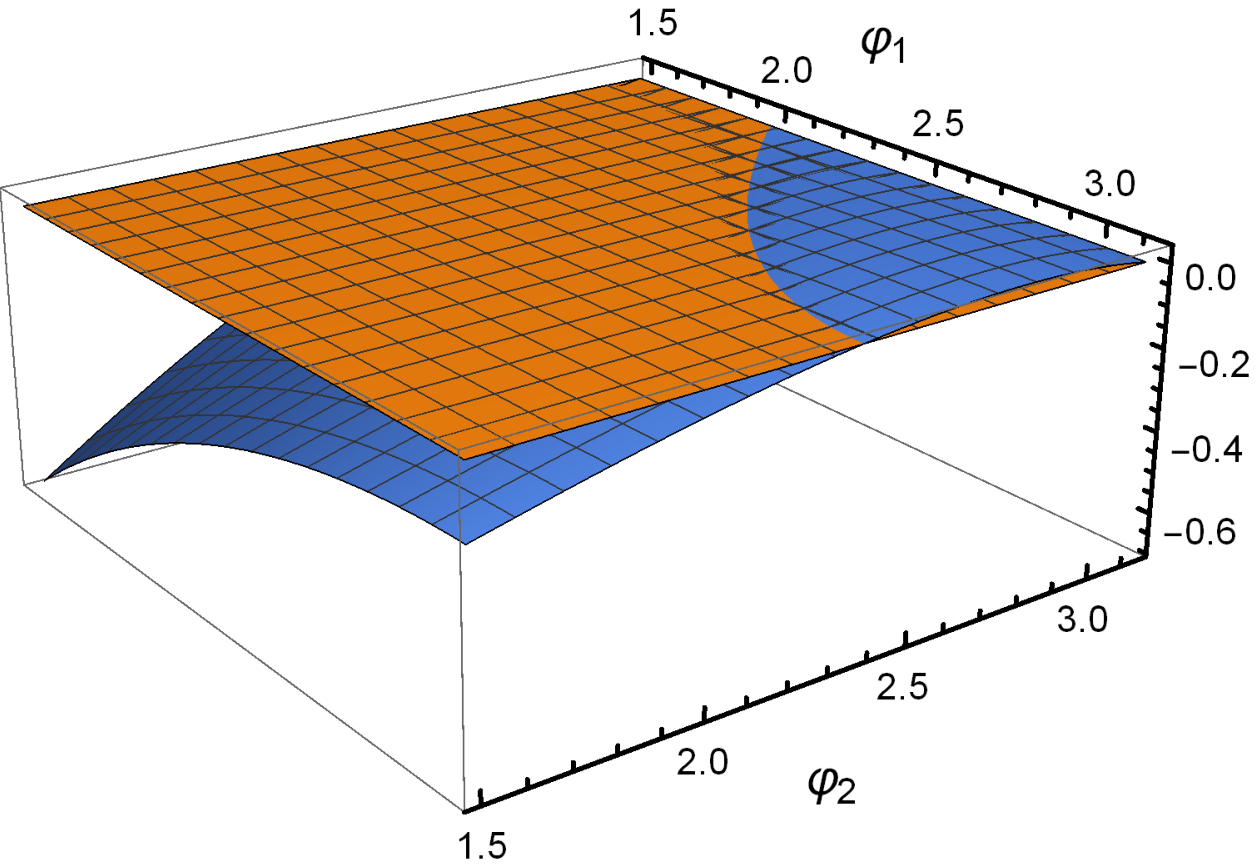}}}
\subfloat[$g^{d1}(\varphi _1, \varphi _2)$]{\resizebox{0.33\textwidth}{!}{\includegraphics{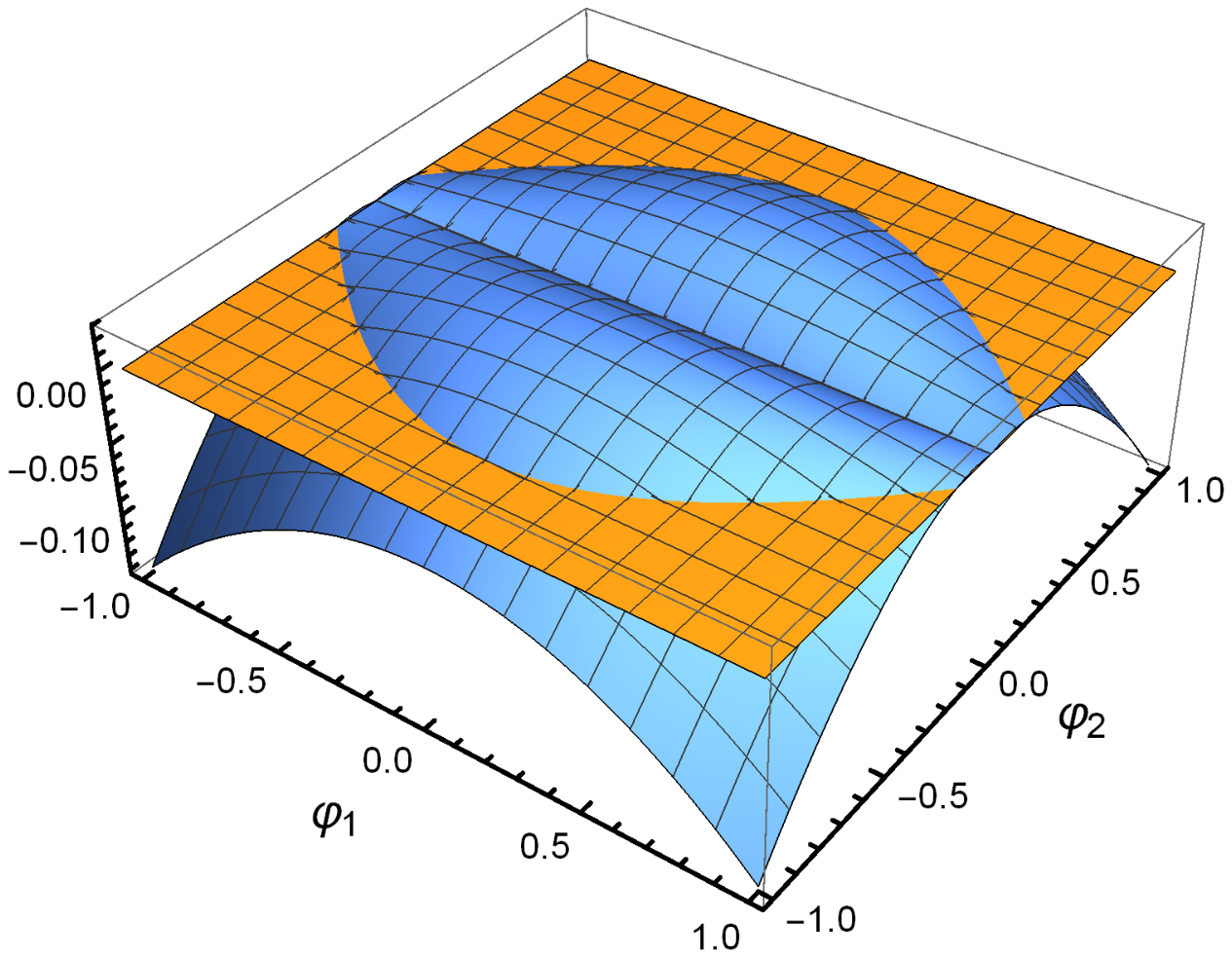}}}
\subfloat[$g^{d2}(\varphi _1, \varphi _2)$]{\resizebox{0.33\textwidth}{!}{\includegraphics{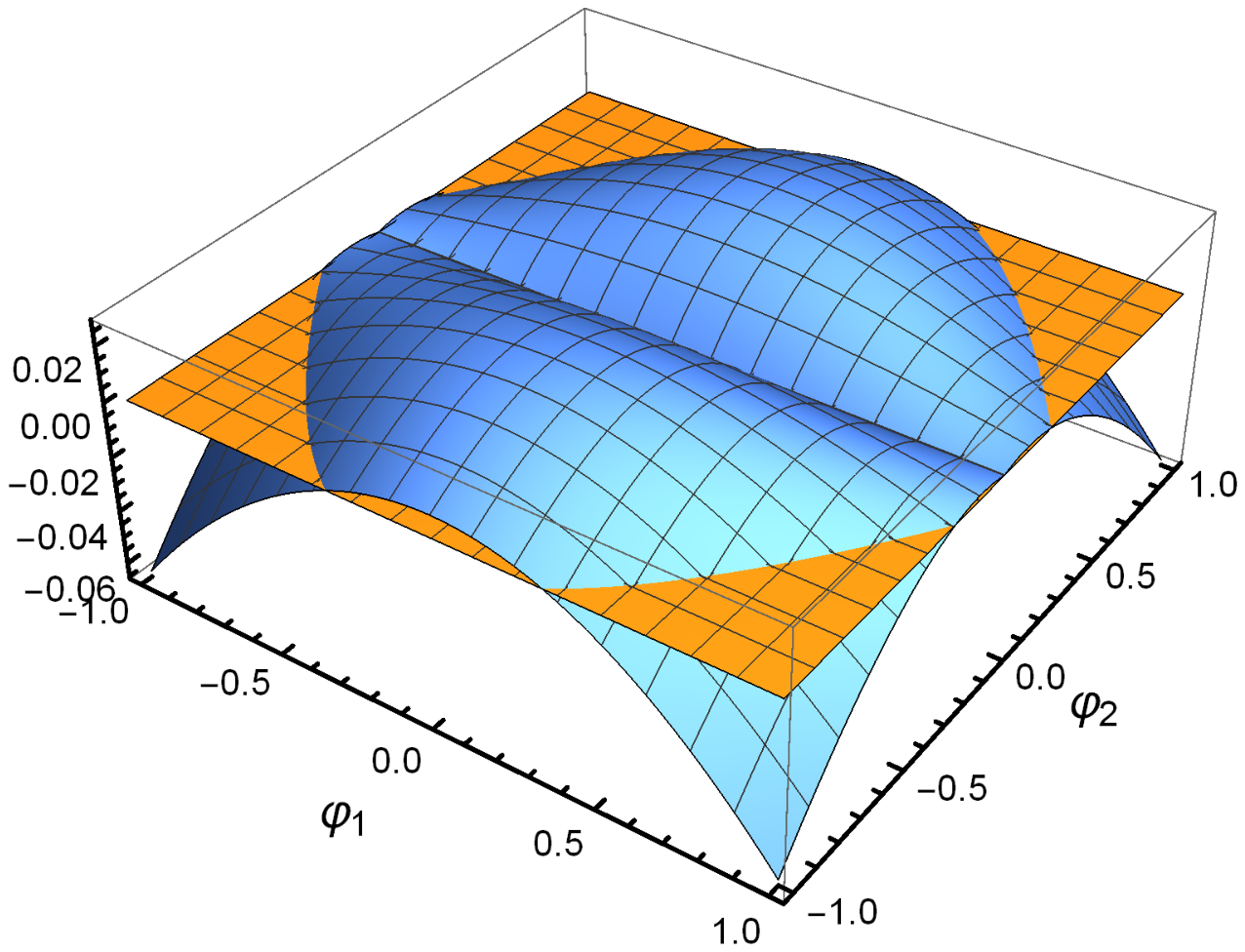}}} \\
\subfloat[$g^u(\pi, \varphi _2)$]{\resizebox{0.33\textwidth}{!}{\includegraphics{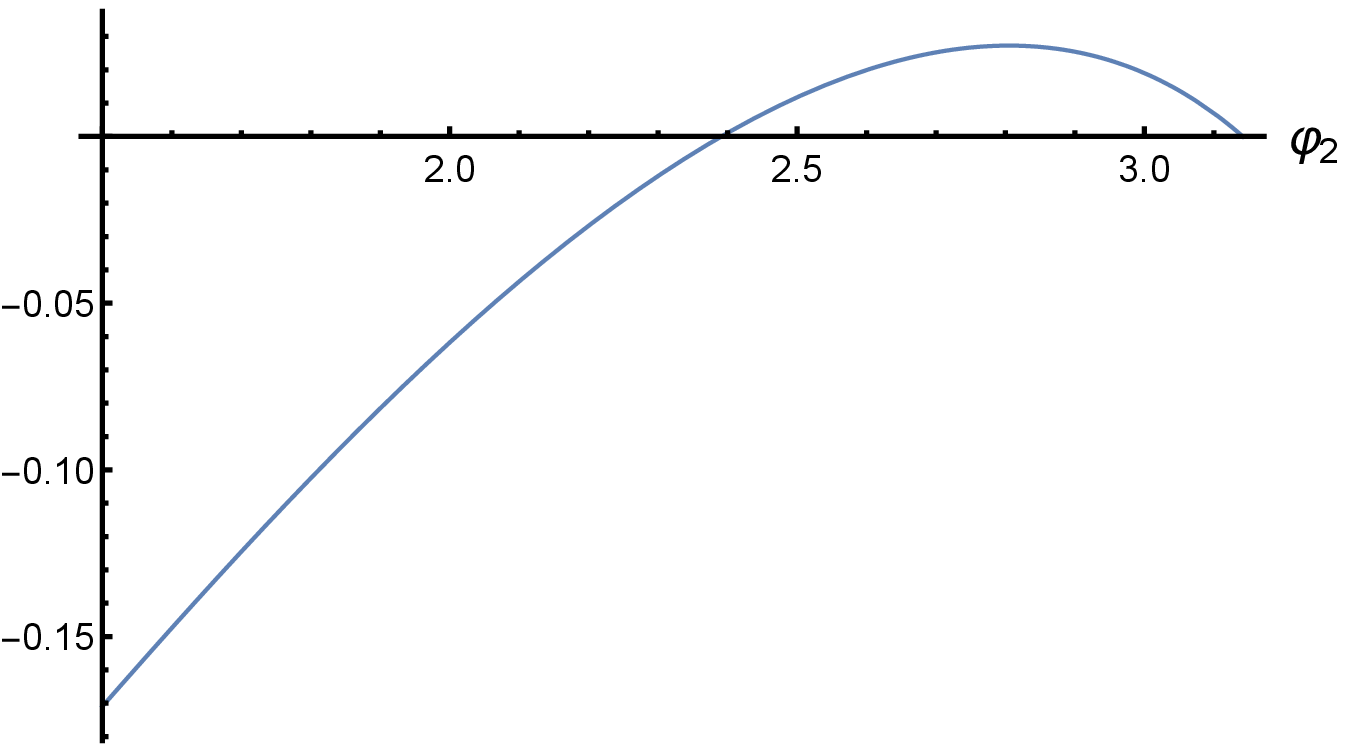}}}
\subfloat[$g^{d1}(0, \varphi _2)$]{\resizebox{0.33\textwidth}{!}{\includegraphics{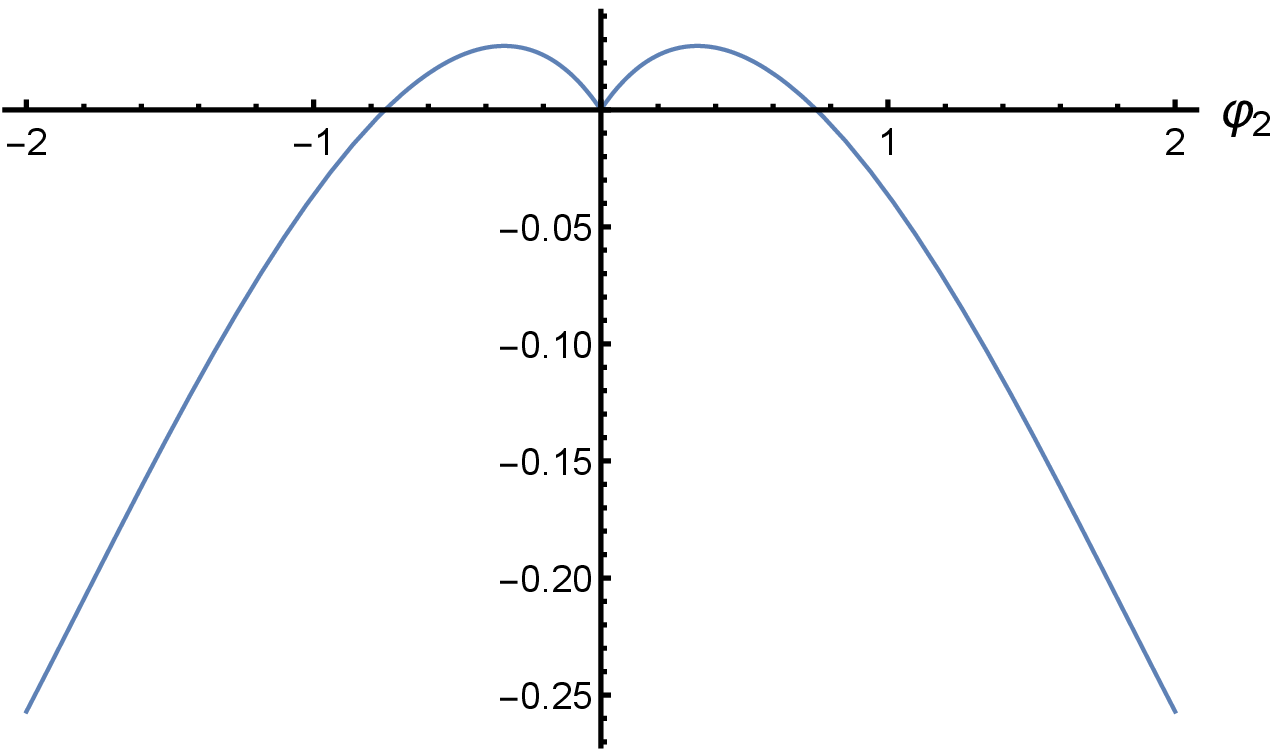}}}
\subfloat[$g^{d2}(0, \varphi _2)$]{\resizebox{0.33\textwidth}{!}{\includegraphics{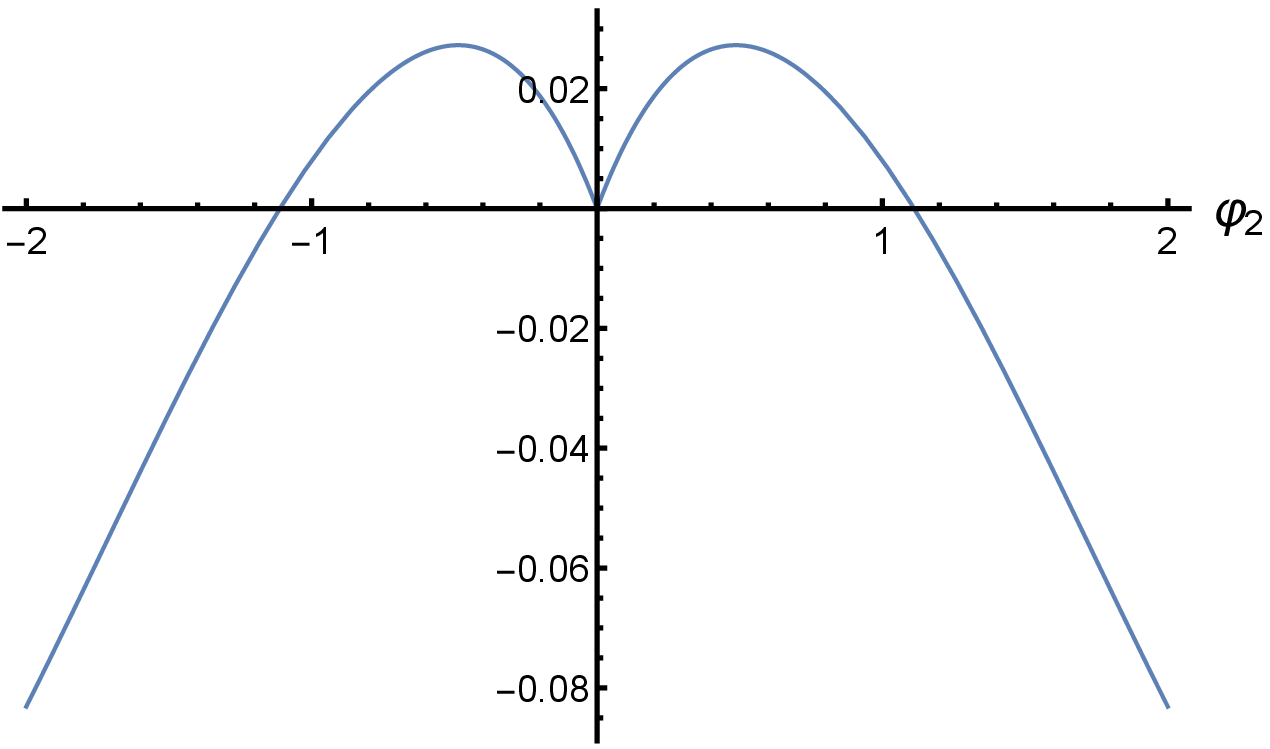}}}\\
\caption{\label{Fig:gud}
LI violation of entangled $B_d$ mesons. $\theta_s$ is fixed to be  $\sin ^{-1}{0.695}$.      Our LI is violated when $g^u(\varphi_1,\varphi_2)>0$  or $g^{d1}(\varphi_1,\varphi_2)>0$ or $g^{d2}(\varphi_1,\varphi_2)>0$.   $g^u>0$  when $\varphi _1=\pi$, $ 2.39<|\varphi _2|<\pi$, and the maximum is at $\varphi _1=\pm \pi, \varphi _2\approx \pm 2.81$.  $g^{d1} >0$ when $\varphi _1=0$,   $0<|\varphi _2|<0.75$, and the maximum is at $\varphi _1=0, \varphi _2\approx \pm 0.336$. $g^{d2} >0$ when $\varphi _1=0$,   $0<|\varphi _2|<1.11$,  and  the maximum is at $\varphi _1=0, \varphi _2\approx \pm 0.486$. The maxima are all about $2.7\%$.}
}
\end{figure*}

Solving $\partial_{\varphi _2}g^{d2}(0,\varphi _2)=0$ numerically, we  find that $g^{d2}(0,\varphi _2)$ reaches its maximum at $\varphi _2\approx \pm 0.486$, and we find numerically  that $g^{d_2}(0,\varphi _2)>0$ when   $ 0<|\varphi _2|<1.11$. The  $\varphi _2$ range   of  $g^{d2}>0$ is larger than that of  $g^{d1}>0$. The maxima of  $g^u$,  $g^{d1}$ and  $g^{d2}$  are all about $2.7\%$. The results are depicted in Fig.~\ref{Fig:gud}.

A $B_d$ meson is unstable, and the time interval between the two decays is of the order of the lifetime $\tau _B$~\cite{TViolationD}. So it is better to study  the case  in which $\Delta t$  is of the  order of $\tau _B$, so that the number of events is large. Thus it is easier to test $g^{d2}$, because  in its  violation region, $\Delta t$ is closer to $\tau _B$.

\begin{figure*}
\centering{
\resizebox{0.9\textwidth}{!}{\includegraphics{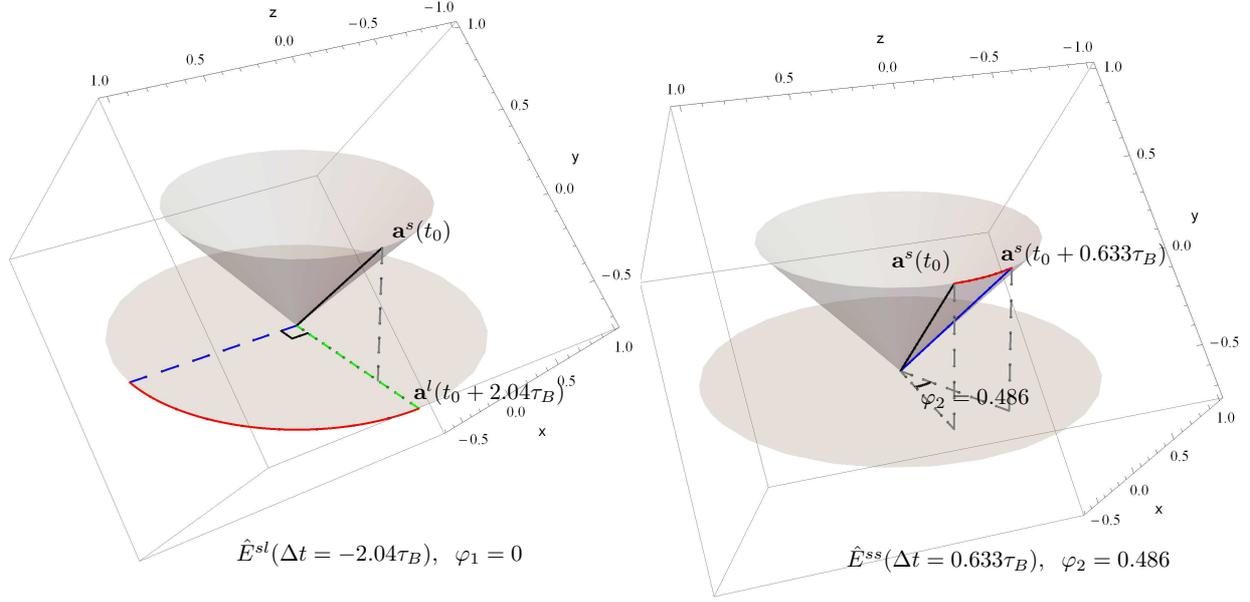}}
\caption{\label{Fig:measure}
The correlation functions to be measured. The left picture represents   $\hat{E}^{sl}(\Delta t = \pi / 2x\Gamma \approx 2.04\tau _B)=\hat{E}^{sl}(\varphi_1=0)$. The right picture represents $\hat{E}^{ss}(\Delta t = 0.486 / x\Gamma \approx 0.633\tau _B)=\hat{E}^{ss}(\varphi _2\approx 0.486)$. They give rise to $g^{d1}( \varphi_1=0,\varphi _2\approx 0.486)$, which maximally violates the second lower bound of our LI.}
}
\end{figure*}

We now focus on how to make measurements to confirm the violation of the second lower bound.  $g^{d2}(\varphi _1,\varphi _2)>0$  can be found   in the regime  $(\varphi _1=0$,  $ 0<|\varphi _2|<1.11)$, as calculated above. The function  $g^{d2}(\varphi _1,\varphi _2)$ contains
$\hat{E}^+_{sl}(\varphi_1)=\hat{E}^{sl}(\varphi_1)+
\hat{E}^{ll}(0)$ and
$\hat{E}^+_{ss}(\varphi_2)=\hat{E}^{ss}(\varphi)+
\hat{E}^{ss}(0)$.
Hence one first  measures $\hat{E}^{ll}(\Delta t_1=0)$ and    $\hat{E}^{sl}(\varphi_1=0)=  \hat{E}^{sl}(\Delta t_1 =-\pi / 2x \Gamma \approx -2.04\tau _B )$,  as shown in Fig.~\ref{Fig:measure}. Thus $\hat{E}^+_{sl}(\varphi_1=0)= \hat{E}^{sl}(\varphi_1=0)+\bar{E}^{ll}(\varphi_1=0)$   is obtained. One also needs to measure  $\hat{E}^{ss}(\Delta t_2 =0)$ and $\hat{E}^{ss}(\Delta t_2)$ with $0<|\Delta t_2|<1.11/x\Gamma \approx 1.44\tau _B$, such that   $\hat{E}^+_{ss}(\varphi_2)$ with     $0<|\varphi _2|<1.11$ is  obtained. Thus one obtains    $g^{d2}(\varphi _1,\varphi _2)>0$ in this regime.   The violation is maximal when $\Delta t_2 \approx 0.486/x\Gamma \approx 0.633\tau _B$, then $g^{d2}(\varphi _1,\varphi _2)\approx
2.7\%$,  as shown in Fig.~\ref{Fig:measure}. Typically, the resolution of the signal is proportional to the inverse of square of event numbers~\cite{TViolationD}. Therefore it  can  be estimated that the expected signal of  LI violation  can be observed when the event number is about $10^4 \sim 10^5$, which can be achieved in  current  experiments~\cite{ImprovedCPV}.

It is also possible to test LI in polarization-entangled baryon-antibaryon pairs produced in, say,  $J/\Psi \to \Lambda \bar{\Lambda}$ decays, where the polarizations of $\Lambda$ and $\bar{\Lambda}$ can be measured through the angular distribution of the momenta of their decay products pions.  However, the effective measuring directions satisfying the Malus' law are yet to be found out.

\begin{figure*}
\centering{
\subfloat[$g^u(\theta_s,\pi, \varphi _2)$]{\resizebox{0.33\textwidth}{!}{\includegraphics{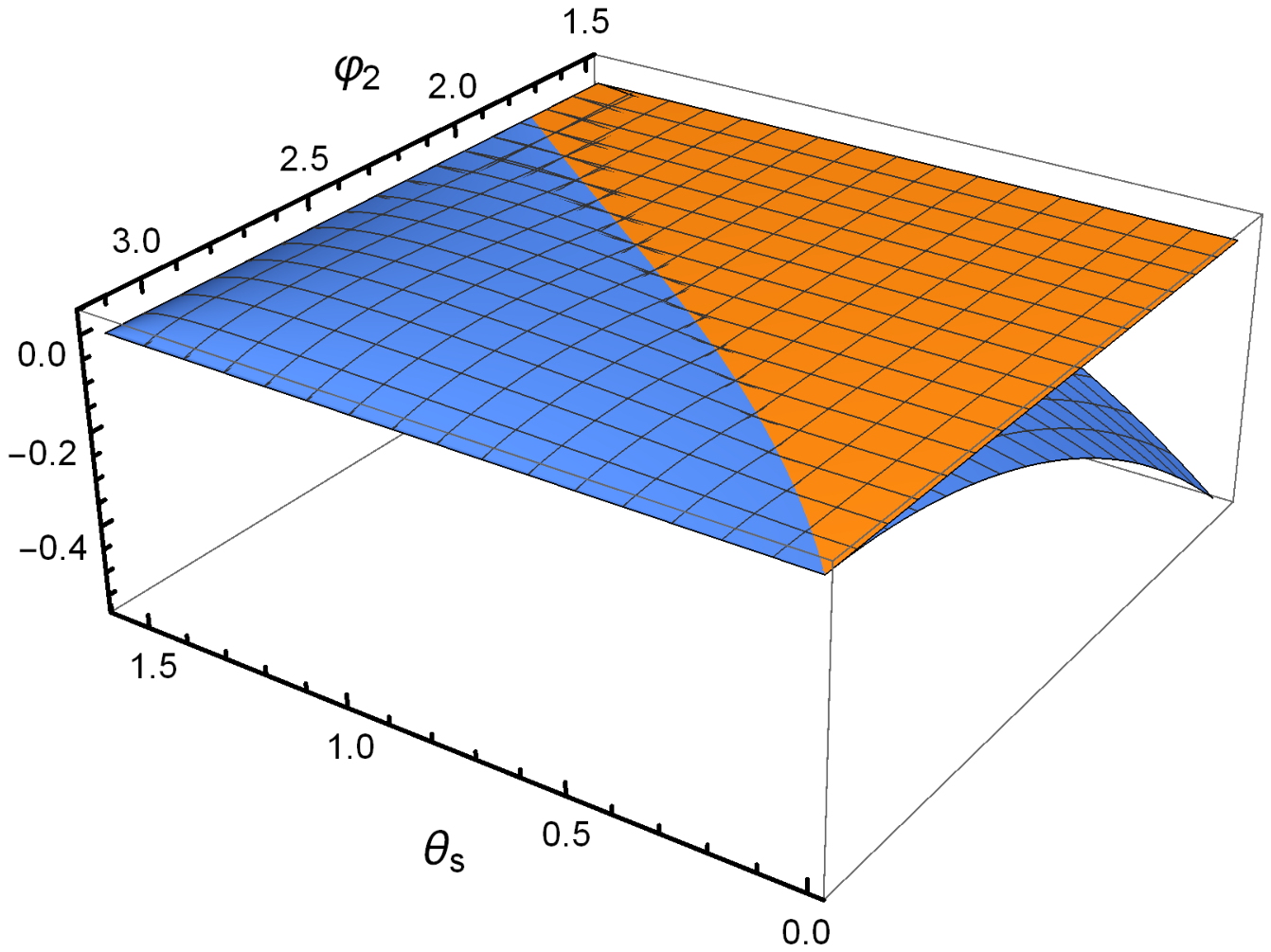}}}
\subfloat[$g^{d1}(\theta_s,0, \varphi _2)$]{\resizebox{0.33\textwidth}{!}{\includegraphics{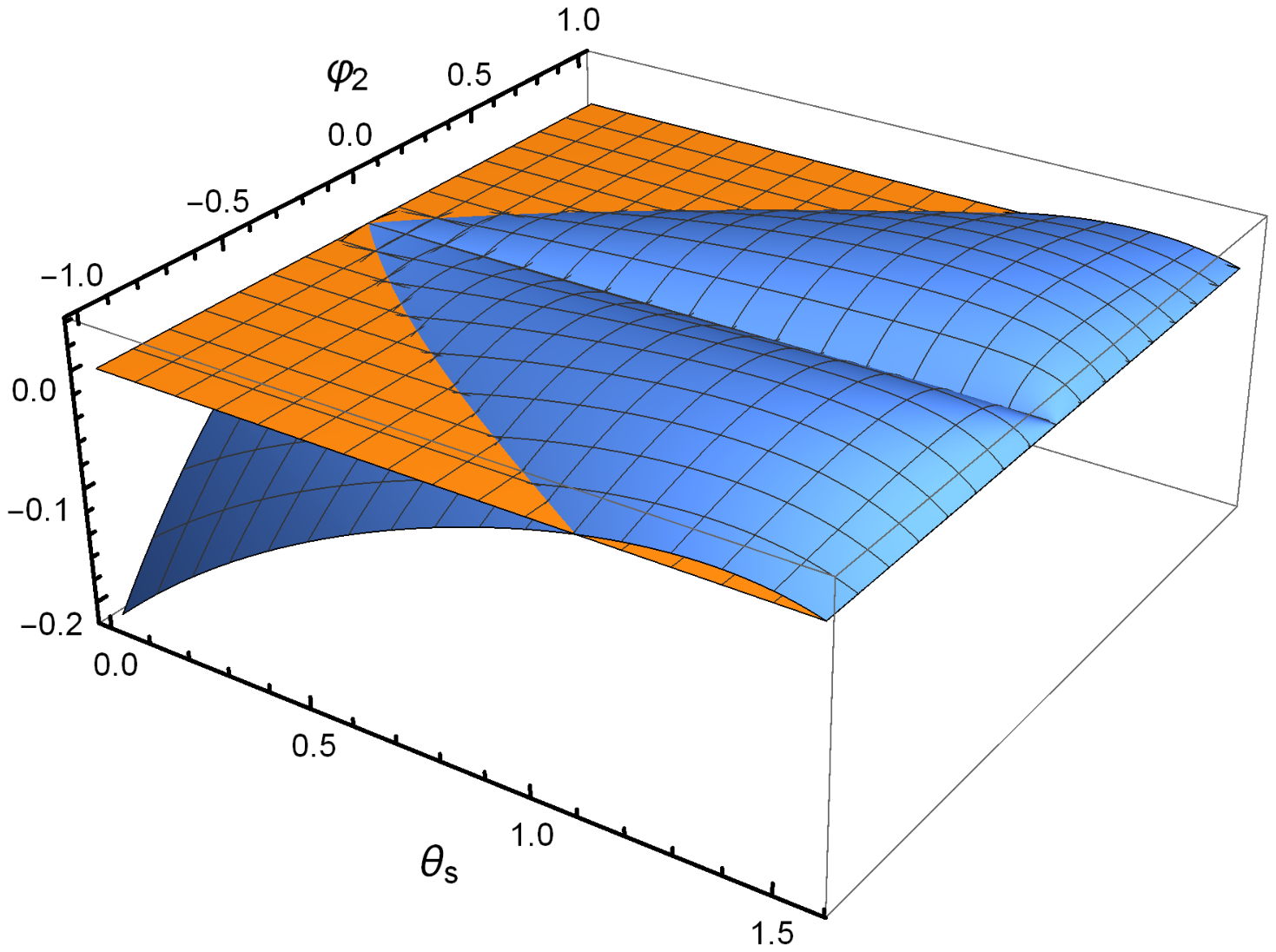}}}
\subfloat[$g^{d2}(\theta_s,0, \varphi _2)$]{\resizebox{0.33\textwidth}{!}{\includegraphics{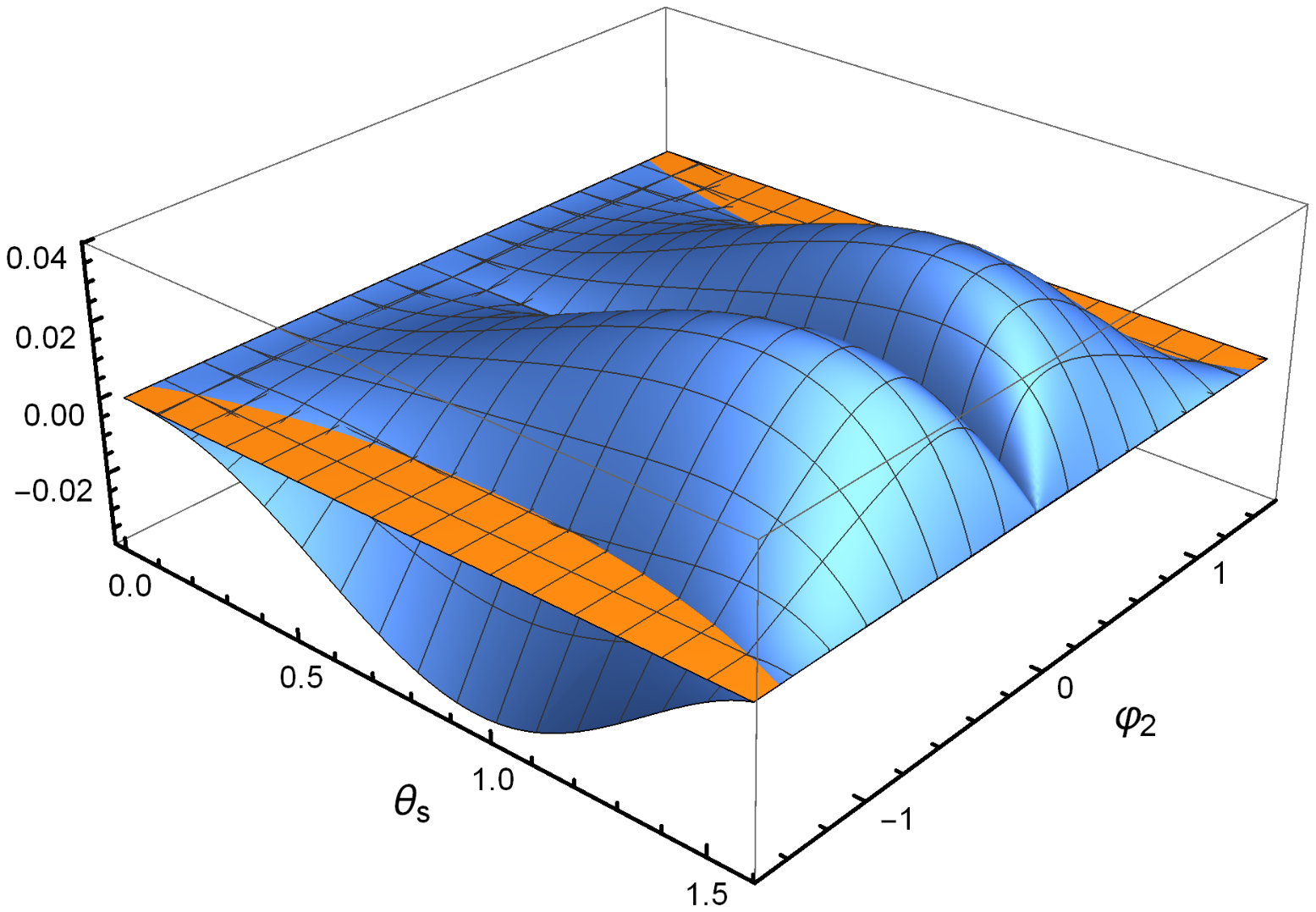}}}
\caption{\label{Fig:g3d}
LI violation in case $\theta_s$ is a variable. The maxima  of $g^u$, $g^{d1}$ and $g^{d2}$ are still at $\varphi _1= \pi,0,0$, respectively.   $g^u(\theta_s,\pi,\varphi _2)$, $g^{d1}(\theta_s,0,\varphi _2)$ and $g^{d2}(\theta_s,0,\varphi _2)$  are functions of $\theta_s$ and $\varphi _2$.
}}
\end{figure*}

If using other systems such as photon polarizations to test our LI, $\theta_s$ may become a variable. The results of $\partial _{\varphi _1}g^u(\varphi _1,\varphi _2)= \partial _{\varphi _1}g^{d1}(\varphi _1,\varphi _2)= \partial _{\varphi _1}g^{d2}(\varphi _1,\varphi _2)=0$ do not depend on $\theta_s$, hence their maxima are still at $\varphi _1= \pi,0,0$, respectively.  $g^u(\theta_s,\pi,\varphi _2)$, $g^{d1}(\theta_s,0,\varphi _2)$ and $g^{d2}(\theta_s,0,\varphi _2)$  as functions of $\theta_s$ and $\varphi _2$  are shown in Fig.~\ref{Fig:g3d}. Interestingly,   in a  certain range of $\varphi _2$, for any value of $\theta _s$ except $0$ and $\pi/2$,  we always have $g^{d2}>0$, that is,  the second lower bound is always violated.

\begin{figure}
\centering
\resizebox{1.0\hsize}{!}{\includegraphics{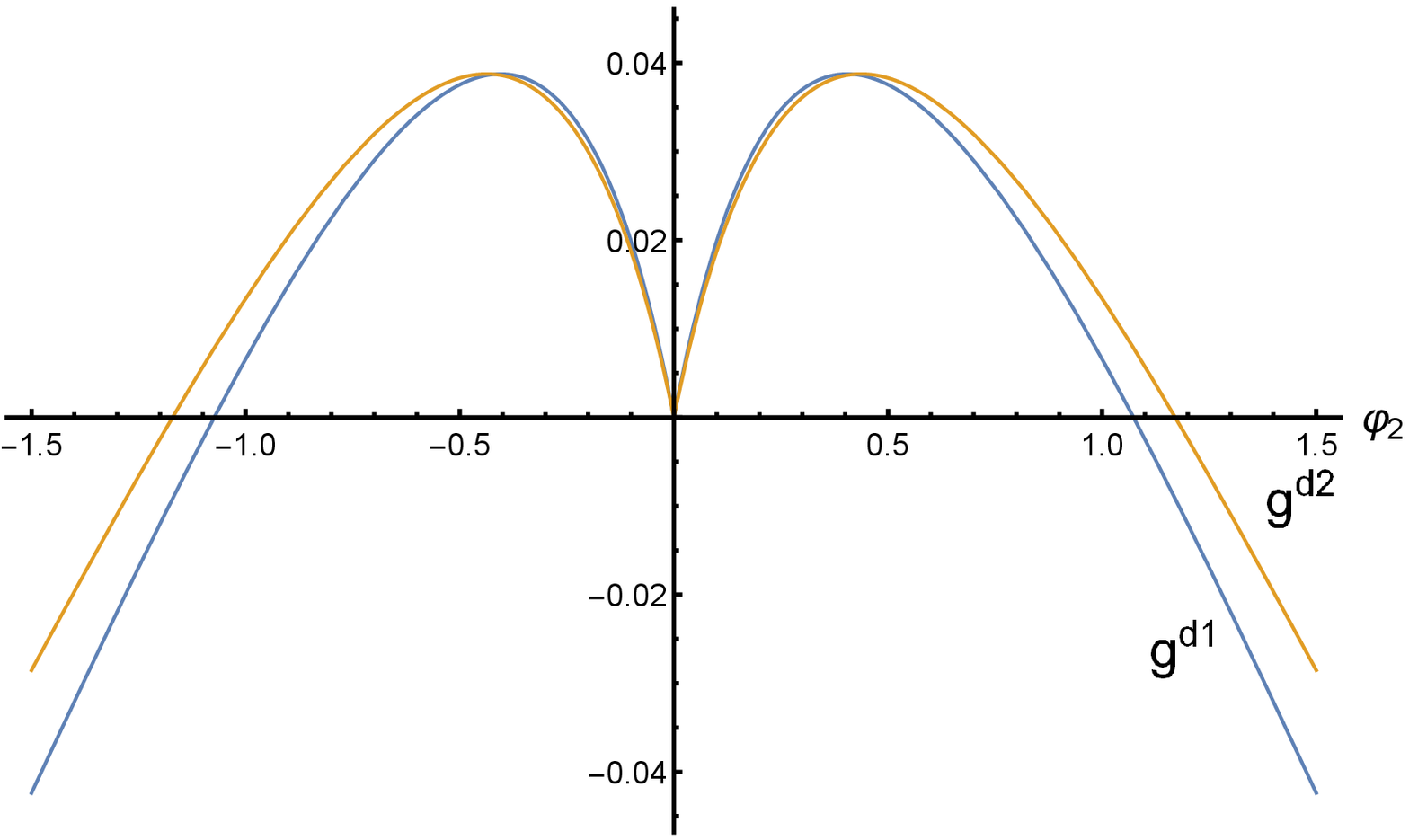}}
\caption{  \label{Fig:gdmax}
The two lower bounds are maximally violated on $\theta_s=1.18208$, $\varphi_1=0$.   Shown here are  the dependence of  $g^{d1}(\theta_s=1.18208,\varphi_1=0,\varphi_2)$ and  $g^{d2}(\theta_s=1.18208,\varphi_1=0,\varphi_2)$  on $\varphi_2$.  $g^{d1} >0$ when   $0<|\varphi_2|<1.0734$, while   $g^{d2}  >0$ when $0<|\varphi_2|<1.17078$. }
\end{figure}

Furthermore, we numerically found the maximal violations  of the three bounds are all $3.87482\%$  when  $\theta _s=1.18208$, i.e.  \begin{equation}\begin{split}
&g^u(1.18208,\pm \pi, \pm 2.7373)=g^{d1}(1.18208,0, \pm 0.404296)\\&=g^{d2}(1.18208,0, \pm 0.437399)=3.87482\%.\end{split}
\end{equation}
It is also found that
\begin{equation}
g^{d1}(1.18208,0,\varphi_2)>0,\,\mbox{when}\,\, 0<|\varphi_2|<1.0734, \end{equation}
\begin{equation} g^{d2}(1.18208,0,\varphi_2)>0,\,\mbox{when}\, \, 0<|\varphi_2|<1.17078.
\end{equation}
The latter is wider, as   shown in Fig~\ref{Fig:gdmax}.

\section{\label{sec:level6} Summary  and   Discussions}

To summarize,  we have extended the CNHV theories to include the case that the measuring settings, together with the measurement outcomes,  are  not externally fixed, but determined by HVs. The outcomes of those measurements with the same settings give averages satisfying Malus' Law and make up correlation functions. We show that such is the case of meson decays,  which  could be determined by HVs at the source of the meson pairs. This extension does not  change  the validity of LI. Therefore, entangled meson pairs can be used to test LI.

We find that for a $B_d$ meson, the effective measuring directions appearing in Malus' Law are  on a cone and a plane, corresponding to semileptonic decays and decays to CP eigenstates, respectively.  For such effective measuring directions, we present a new  LI.  This can be tested in   $C=-1$ entangled  state of $B^0-\bar{B}^0$ pairs, within the present experimental capability. The expected violation is estimated quantitatively, using the indirect CP violation and other parameters.  Our LI is violated  if there is indirect CP violation. There may be  profound reason for this surprising connection. Besides, our new LI can also be tested in other systems such as photon polarizations, where the measuring directions are simply directions externally fixed.

\begin{acknowledgement}
This work is supported by  National Science Foundation of China (Grant No. 11574054 and No. 11947066).
\end{acknowledgement}

\appendix

\section{Two Inequalities}

We derive a new LI, for ${\bf a}^l(\phi_l)$ on a plane and ${\bf a}^s(\theta_s,\phi_s)$ on a cone.

For each particle,  each measurement  yields an outcome $A=A({\bf u},{\bf v},\tilde{\bf a}(\lambda),\tilde{\bf b}(\lambda),\lambda)=\pm 1$ or $B= B({\bf v},{\bf u},\tilde{\bf b}(\lambda),\tilde{\bf a}(\lambda),\lambda)=\pm 1$.   Using
$-1+\int d\lambda \rho' _{{\bf u},{\bf v},{\bf a},{\bf b}}(\lambda) \left|A+B\right|=\int d\lambda \rho'_{{\bf u},{\bf v},{\bf a},{\bf b}}(\lambda)AB=1-\int d\lambda \rho' _{{\bf u},{\bf v},{\bf a},{\bf b}}(\lambda)\left|A-B\right|$,   where $\rho'_{\mathbf{u},\mathbf{v},\mathbf{a},\mathbf{b}}(\lambda) \equiv  \rho_{\mathbf{u},\mathbf{v},\mathbf{v}} (\lambda)  \delta(\tilde{\bf a}(\lambda)-\mathbf{a})\delta(\tilde{\bf b}(\lambda)-\mathbf{b})$,
and  $\bar{A}=\int d \lambda \rho'_{\mathbf{u},\mathbf{v},\mathbf{a},\mathbf{b} }(\lambda)   A(\mathbf{u},\mathbf{v},\tilde{\bf a}(\lambda),\tilde{\bf b}(\lambda),\lambda)
 =\mathbf{u}\cdot \mathbf{a}$, $\bar{B}=\rho'_{\mathbf{u},\mathbf{v},\mathbf{a},\mathbf{b} }(\lambda)    B(\mathbf{v},\mathbf{u},\tilde{\bf b}(\lambda),\tilde{\bf a}(\lambda),\lambda)
 =\mathbf{v}\cdot \mathbf{b}$, one  finds $1-\int d{\bf u}d{\bf v}F({\bf u},{\bf v})|{\bf u}\cdot {\bf a}-{\bf v}\cdot {\bf b}| \geq E({\bf a},{\bf b}) \geq -1+\int d{\bf u}d{\bf v}F({\bf u},{\bf v})|{\bf u}\cdot {\bf a}+{\bf v}\cdot {\bf b}|$~\cite{LI,LIExpNature}.

Furthermore~\cite{comment}, considering
$|{\bf u}\cdot {\bf a}+{\bf v}\cdot {\bf b}|
+|{\bf u}\cdot {\bf b}+{\bf v}\cdot {\bf b}|
\geq \left|{\bf u}\cdot {\bf a}+{\bf v}\cdot {\bf b}-\left({\bf u}\cdot {\bf b}+{\bf v}\cdot {\bf b}\right)\right|
=\left|{\bf u}\cdot \left({\bf a}- {\bf b}\right)\right|$,  and
$|{\bf u}\cdot {\bf a}-{\bf v}\cdot {\bf b}|+
|{\bf u}\cdot {\bf b}+{\bf v}\cdot {\bf b}|
\geq \left|{\bf u}\cdot \left({\bf a} + {\bf b}\right)\right|$, one obtains
\begin{subequations}
\begin{eqnarray}
&\hat{E}^+({\bf a},{\bf b})\geq -2+\int d{\bf u}F({\bf u})|{\bf u}\cdot \left({\bf a}-{\bf b}\right)|, \label{eq.3.7a}\\
&\hat{E}^-({\bf a},{\bf b})\leq 2-\int d{\bf u}F({\bf u})|{\bf u}\cdot \left({\bf a}+ {\bf b}\right)|. \label{eq.3.7b}
\end{eqnarray}
\label{eq.3.7}
\end{subequations}
with $\hat{E}^+({\bf a},{\bf b})\equiv E({\bf a},{\bf b} )+ E({\bf b},{\bf b})$ and $\hat{E}^-({\bf a},{\bf b})\equiv E({\bf a},{\bf b} )+ E({\bf b},-{\bf b})$.

All the results remain valid in the special case that $\tilde{\bf a}(\lambda)$ and  $\tilde{\bf b}(\lambda)$ are externally set to be always  ${\bf a}$ and  $\mathbf{b}$ respectively.

\section{  Upper bound}

\begin{figure*}
\centering
\resizebox{0.8\textwidth}{!}{\includegraphics{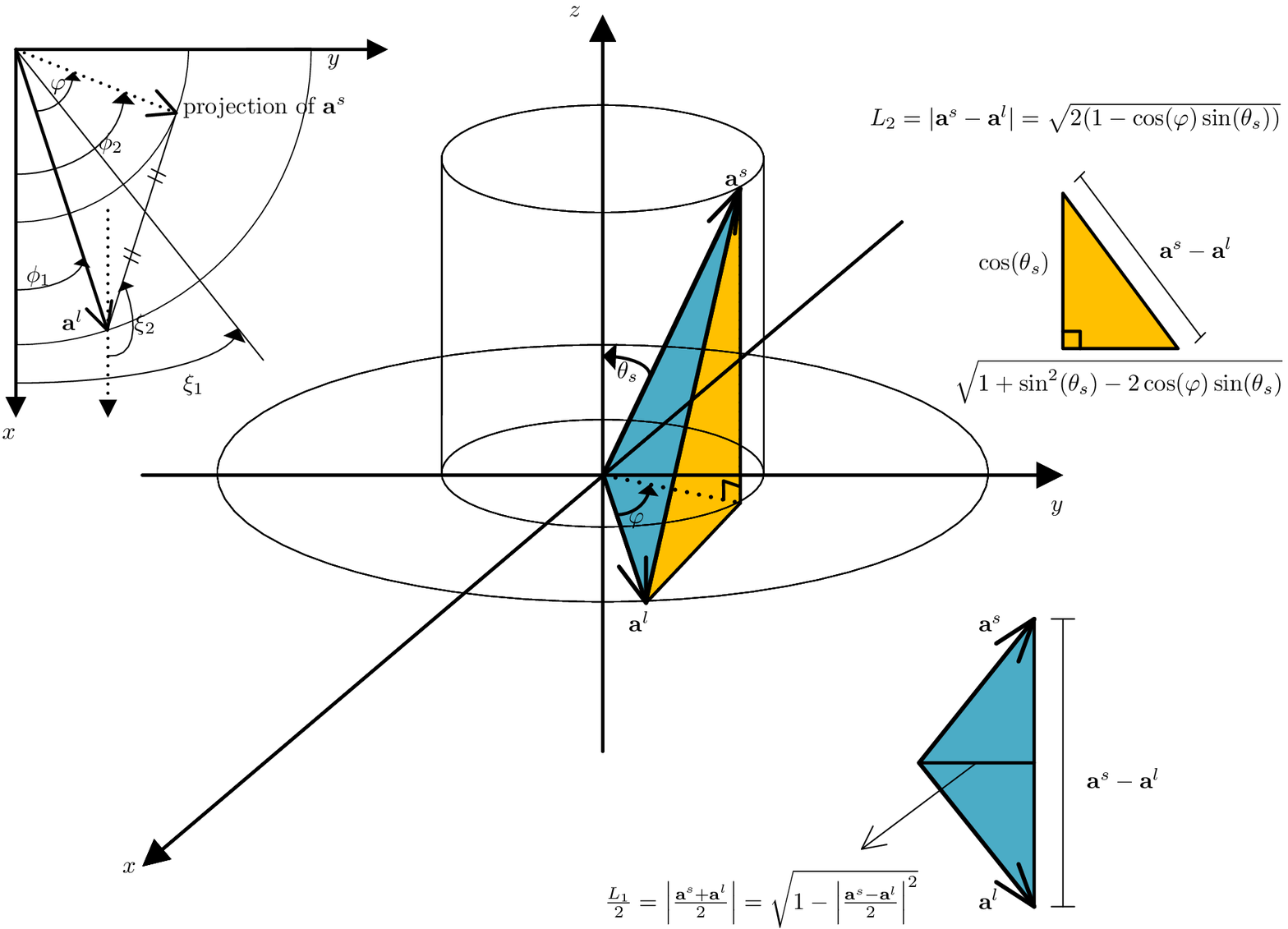}}
\caption{\label{Fig:fig2} Geometric relations among ${\bf a}^s$, ${\bf a}^l$, ${\bf a}^s+{\bf a}^l$ and ${\bf a}^s-{\bf a}^l$.}
\end{figure*}

Suppose  ${\bf a} = {\bf a}^s(\theta_s,\phi_a)$,   ${\bf b}={\bf a}^l(\phi_b)$. Then
Eqs.~(\ref{eq.3.7b}) reads
\begin{equation}
\begin{split}
&\hat{E}^{-}({\bf a}^s(\theta_s,\phi_a),{\bf a}^l(\phi_b))  \leq 2-\int _0^{\pi}\sin \theta _ud\theta _u \\
&\times \int _0^{2\pi} d\phi _uF(\theta _u,\phi _u)\left|{\bf u}(\theta _u,\phi _u)\cdot ({\bf a}^s(\phi _a)+ {\bf a}^l(\phi _b))\right|.
\end{split}
\label{eq.3.11}
\end{equation}
As shown in Fig.~\ref{Fig:fig2},
\begin{equation}
\begin{split}
&{\bf a}^s(\theta_s,\phi_a)+{\bf a}^l(\phi_b)=\\
&L_1(\theta_s,\varphi) \left(\sin(\theta _{1})\cos (\xi+\alpha_1),\sin (\theta _{1})\sin (\xi _1),\cos (\theta _{1})\right),\\
\end{split}
\label{eq.3.12}
\end{equation}
where $$L_1(\theta_s,\varphi)\equiv |{\bf a}^s+{\bf  a}^l|=\sqrt{2+2\cos(\varphi)\sin (\theta _s)},$$ $$\theta _{1}(\theta_s,\varphi)=\cos ^{-1}\frac{\cos (\theta _s)}{\sqrt{2+2\cos (\varphi)\sin (\theta _s)}},$$  $\varphi = \phi _a-\phi _b,$  $\xi =\frac{\phi_a+\phi_b}{2}$, $\alpha _1$ is an angle depending on $\varphi$ and $\theta _s $ while  independent of $\xi$, as shown in Fig.~\ref{Fig:fig2}.  With $0<\theta _s<\pi /2$, we have  $\sin(\theta _{1})> 0$, $\cos(\theta _{1})> 0$.

Rewriting  $  \hat{E}^{-}({\bf a}^s(\theta_s,\phi_a),{\bf a}^l(\phi_b)) $  as $ \hat{E}^{-}_{sl}(\theta _s,\xi,\varphi)$, then
\begin{equation}
\begin{split}
&\hat{E}^{-}_{sl} (\theta _s,\varphi)\equiv \int \frac{d\xi }{2\pi}  \hat{E}^{-}_{sl}  (\theta _s,\xi,\varphi) \\
&\leq  2 -L_1(\theta_s,\varphi)\int _0^{2\pi}\frac{d\xi}{2\pi}\int _0^{\pi}\sin \theta _ud\theta _u \int _0^{2\pi} d\phi _uF(\theta _u,\phi _u)\\
&\times \left|\sin (\theta _u) \sin (\theta _{1}) \cos (\phi _u-\xi-\alpha_1)+\cos (\theta _u) \cos (\theta _{1})\right|\\
&\leq  2 -L_1(\theta_s,\varphi) \int _0^{\pi}\sin \theta _ud\theta _u F(\theta _u)\cos (\theta _{1})\left|\cos (\theta _u)\right|,\\
\end{split}
\label{eq.3.17}
\end{equation}
where $F(\theta _u)\equiv \int _0^{2\pi} d\phi _u F(\theta _u,\phi _u)$. In obtaining the second inequality, we have used
\begin{equation}
\begin{split}
&\int _0^{2\pi}\frac{d\xi }{2\pi}\left|\sin (\theta _{1}) \cos (\phi _u-\xi-\alpha_1 )\sin (\theta _u)+\cos (\theta _{1})\cos (\theta _u)\right|\\
&=\int _0^{2\pi}\frac{d\xi }{2\pi}\left|\sin (\theta _{1}) \cos (\xi)\sin (\theta _u)+\cos (\theta _{1})\cos (\theta _u)\right|  \\
&\geq \cos (\theta _{1})\left|\cos (\theta _u)\right|,
\end{split}
\label{eq.3.16}
\end{equation}
as
\begin{equation*}
\begin{split}
&\int _0^{2\pi}d\xi \left|\cos (\xi + \beta)+a\right|=\int _0^{2\pi}d\xi \left|\cos (\xi)+a\right|\\
&=4{\rm Re}\left(\sqrt{1-a^2}+a\sin^{-1}(a)\right)\geq 2\pi |a|,
\end{split}
\end{equation*}
where  $a$  and $\beta$ are arbitrary real numbers.

The case that the two vectors are on a same plane is a special case of above with  $\theta_s=\pi / 2$. In this special case, $L_1=2\cos \left(\varphi /2\right)$,  $\theta _{1}=\pi / 2$. Hence
\begin{equation}
\begin{split}
&\hat{E}^{-}_{ll}(\theta _s,\varphi)\equiv \int \frac{d\xi}{2\pi}E({\bf a}^l(\xi + \frac{\varphi}{2}),{\bf a}^l(\xi -\frac{\varphi}{2}))\\
&\leq 2-2\cos(\frac{\varphi }{2})\int _0^{2\pi}\frac{d\xi}{2\pi}\int _0^{\pi}\sin \theta _ud\theta _u F(\theta _u)\left|\cos (\xi)\sin (\theta _u)\right|\\
&= 2 -\frac{4}{\pi}\cos (\frac{\varphi '}{2})\int _0^{\pi}\sin \theta _ud\theta _u F(\theta _u)\left|\sin (\theta _u)\right|,\\
\end{split}
\label{eq.3.18}
\end{equation}
where $\int _0^{2\pi}d\xi |\cos(\xi)|=4$ is used.

Therefore one obtains
\begin{equation}
\begin{split}
&\hat{E}^{-}_{sl}(\theta _s,\varphi _1)+\frac{\pi \cos (\theta _{1}(\theta _s, \varphi _1)) L_1(\theta_s,\varphi _1)}{4\cos (\frac{\varphi _2}{2})}{\hat{E}^{ll-}}(\theta _s,\varphi _2)\\
&\leq  2\left(1+\frac{\pi \cos (\theta _{1}(\theta _s, \varphi _1)) L_1(\theta_s,\varphi _1) }{4\cos (\frac{\varphi _2}{2})}\right)\\
&
-\cos (\theta _{1}(\theta _s, \varphi _1)) L_1(\theta_s,\varphi _1)\\
&\times \int _0^{\pi}\sin \theta _ud\theta _u F(\theta _u)\left(|\cos (\theta _u)|+|\sin (\theta _u)|\right).\\
\end{split}
\label{eq.3.21}
\end{equation}

Using $\int _0^{\pi}d\theta _u \sin (\theta _u) F(\theta _u)=1$ and $|\cos (\theta _u)|+|\sin (\theta _u)| \geq 1  $,   one obtains
\begin{equation}
\begin{split}
& \hat{E}^{-}_{sl} (\theta _s,\varphi _1)+\frac{\pi \cos (\theta _{1}(\theta _s, \varphi _1)) L_1(\theta_s,\varphi _1) }{4\cos(\frac{\varphi _2}{2})} \hat{E}^{-}_{ll}(\theta _s,\varphi _2)  \\ & \leq 2  \left(1+\frac{\pi \cos (\theta _{1}(\theta _s, \varphi _1)) L_1(\theta_s,\varphi _1)  }{4\cos(\frac{\varphi _2}{2})}\right)\\
&-\cos (\theta _{1}(\theta _s, \varphi _1)) L_1(\theta_s,\varphi _1).
\end{split}
\label{eq.3.25}
\end{equation}

\section{Lower bounds}

From Eq.~(\ref{eq.3.7a}), one obtains
\begin{equation}
\begin{split}
& \hat{E}^{+} ({\bf a}^s(\theta _s,\phi _a),{\bf a}^l(\phi _b))\geq -2+\int _0^{\pi}\sin \theta _ud\theta _u \int _0^{2\pi} d\phi _u\\
&\times F(\theta _u,\phi _u)\left|{\bf u}(\theta _u,\phi _u)\cdot ({\bf a}^s(\theta _s,\phi_a)-{\bf a}^l(\phi _b))\right|.
\end{split}
\label{eq.3.26}
\end{equation}
As shown in Fig.~\ref{Fig:fig2},
\begin{equation}
\begin{split}
&{\bf a}^s(\theta _s,\phi _a)-{\bf a}^l(\phi _b)=\\
&L_2(\theta_s,\varphi)\left(\sin(\theta _{2})\cos (\xi+\alpha_2),\sin (\theta _{2})\sin (\xi_2),\cos (\theta _{2})\right),
\end{split}
\label{eq.3.27}
\end{equation}
where
\begin{equation*}
\begin{split}
&L_2(\theta_s,\varphi)\equiv |{\bf a}^s-{\bf a}^l|=\sqrt{2-2\cos(\varphi)\sin (\theta _s)}\\
&\theta _{2}=\cos ^{-1}\frac{\cos (\theta _s)}{\sqrt{2-2\cos (\varphi)\sin (\theta _s)}},
\end{split}
\end{equation*}
$\alpha _2$ is an angle depending on $\theta _s$ and  $\varphi$ while independent of $\xi$.

Rewriting  $  \hat{E}^{+}({\bf a}^s(\theta _s,\phi_a),{\bf a}^l(\phi_b)) $  as $ \hat{E}^{+}_{sl}(\theta _s,\xi,\varphi)$,
\begin{equation}
\begin{split}
&\hat{E}^{+}_{sl}  (\theta _s,\varphi)\equiv \int \frac{d\xi }{2\pi}  \hat{E}^{+}_{sl}  (\theta _s,\xi,\varphi)\\
&\geq -2+L_2(\theta_s,\varphi)\int _0^{2\pi}\frac{d\xi}{2\pi}\int _0^{\pi}\sin \theta _ud\theta _u \int _0^{2\pi} d\phi _uF(\theta _u,\phi _u)\\
&\times \left|\sin (\theta _u) \sin (\theta _{2}) \cos (\phi _u-\xi-\alpha_2)+\cos (\theta _u) \cos (\theta _{2})\right|.
\end{split}
\label{eq.3.27.b}
\end{equation}

Considering  the special case  of $\theta _s=\pi / 2$, one obtains
\begin{equation}
\begin{split}
&\hat{E}^+_{ll}(\varphi)\geq -2+2\left|\sin(\frac{\varphi}{2})\right|\int _0^{2\pi}\frac{d\xi}{2\pi}\int _0^{\pi}\sin \theta _ud\theta _u \\
&\times F(\theta _u)\left|\cos(\xi)\sin (\theta _u)\right|.\\
\end{split}
\label{eq.3.28}
\end{equation}
Therefore
\begin{equation}
\begin{split}
&\hat{E}^+_{sl}(\theta _s,\varphi _1)+\frac{\pi \cos (\theta _{2}(\theta_s,\varphi _1))L_2(\theta_s,\varphi _1)}{4\left|\sin(\frac{\varphi _2}{2})\right|}\hat{E}^+_{ll}(\theta_s,\varphi _2)\\
& \geq -2\left(1+\frac{\pi \cos (\theta _{2}(\theta_s,\varphi _1)) L_2(\theta_s,\varphi _1)}{4\left|\sin(\frac{\varphi _2}{2})\right|}\right)\\
&+ \cos (\theta _{2}(\theta_s,\varphi _1)) L_2(\theta_s,\varphi _1).
\end{split}
\label{eq.3.29}
\end{equation}

\begin{figure*}
\centering
\resizebox{0.8\textwidth}{!}{\includegraphics{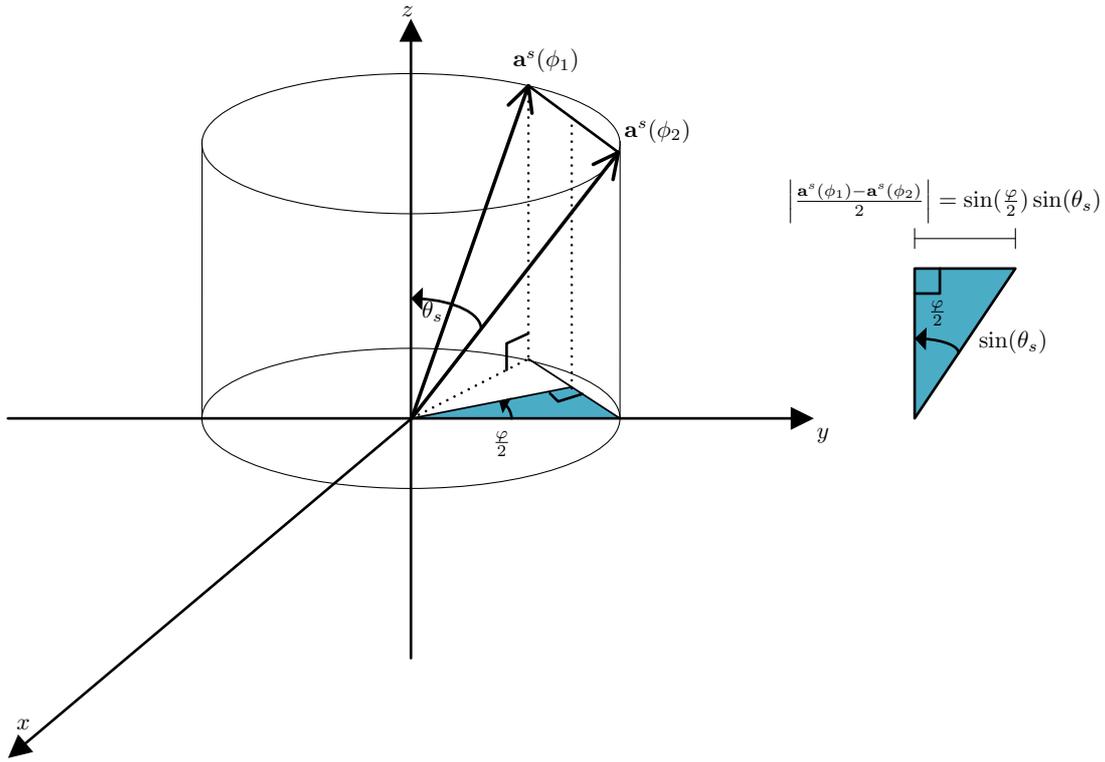}}
\caption{\label{Fig:fig3} Geometric relations of  ${\bf a}^s(\phi _a)$, ${\bf a}^s(\phi _b)$ and ${\bf a}^s(\phi _a)-{\bf a}^s(\phi _b)$.}
\end{figure*}

We find the second lower bound,    in terms of  correlation function  $\hat{E}^{ss+}(\theta_s,\phi _a,\phi _b)$ between ${\bf a}^s(\theta_s,\phi _a)$ and ${\bf a}^s(\theta_s,\phi _b)$, which are on a same  plane, with $|{\bf a}^s(\theta_s,\phi _a)-{\bf a}^s(\theta_s,\phi _b)|=2\sin (\theta _s)\sin (\varphi / 2)$, as shown in Fig.~\ref{Fig:fig3}. We find
\begin{equation}
\begin{split}
&\hat{E}^+_{ss} (\theta_s,\varphi)\geq -2+2\sin(\theta _s)\left|\sin(\frac{\varphi }{2})\right|\\
&\times \int _0^{2\pi}\frac{d\xi}{2\pi}\int _0^{\pi}\sin \theta _ud\theta _u F(\theta _u)\left|\cos(\xi)\sin (\theta _u)\right|.\\
\end{split}
\label{eq.3.30}
\end{equation}

Therefore,
\begin{equation}
\begin{split}
& \hat{E}^+_{sl}(\theta_s,\varphi _1)+\frac{\pi \cos (\theta _2(\theta_s,\varphi _1)) L_2(\theta_s,\varphi_1)}{4\sin(\theta _s)\left|\sin(\frac{\varphi _2}{2})\right|}\hat{E}^+_{ss}(\varphi _2) \\
& \geq -2\left(1+\frac{\pi \cos (\theta _2(\theta_s,\varphi _1)) L_2(\theta_s,\varphi_1)}{4\sin(\theta _s)\left|\sin(\frac{\varphi _2}{2})\right|}\right)\\
&+\cos (\theta _2)L_2(\theta_s,\varphi_1).
\end{split}
\label{eq.3.31}
\end{equation}

Eqs.~(\ref{eq.3.25}), (\ref{eq.3.29}) and (\ref{eq.3.31}) comprise  our LI.

\section{  LI in terms of discrete versions of average correlation functions }

In the discrete version, Eq.~(\ref{eq.3.18}) is  changed to
\begin{equation}
\begin{split}
&\hat{E}^{-}_{sl } (\theta_s,\varphi)\equiv \frac{1}{N}\sum _{n=1}^N{\hat{E}^{-}_{sl}} (\theta_s,\frac{2n\pi}{N},\varphi) \\
&\leq 2-L_1(\theta_s,\varphi)\int _0^{\pi}\sin \theta _ud\theta _u \int _0^{2\pi} d\phi _uF(\theta _u,\phi _u)\\
&\times \frac{1}{N}\sum _{n=1}^N\left|\sin (\theta _u) \sin (\theta _{1} (\theta_s,\varphi)) \cos (\phi _u-\alpha_1-\frac{2n\pi}{N})\right.\\
&\left.+\cos (\theta _u) \cos (\theta _{1} (\theta_s,\varphi))\right|.
\end{split}
\label{eq.3.44}
\end{equation}

Noting
\begin{equation}
\begin{split}
&\frac{1}{N}\sum _{n=1}^N \left|\cos(\frac{2n\pi}{N} + \beta)+a\right|\geq |a|\\
\end{split}
\label{eq.3.48}
\end{equation}
for arbitrary real numbers $\beta$ and $a$, we can change  Eq.~(\ref{eq.3.16})   to
\begin{equation}
\begin{split}
&\frac{1}{N}\sum _{n=1}^N\left|\sin (\theta _{1} (\theta_s,\varphi)) \cos (\phi _u-\frac{2n\pi}{N}-\alpha_1 )\sin (\theta _u)\right.\\
&\left.+\cos (\theta _{1})\cos (\theta _u)\right| \geq \cos (\theta _{1} (\theta_s,\varphi))\left|\cos (\theta _u)\right|.
\end{split}
\label{eq.3.49}
\end{equation}
which is then used in Eq.~(\ref{eq.3.44}). One finds
\begin{equation}
\begin{split}
&\hat{E}^{-}_{sl  } (\theta_s,\varphi)\leq\\
& 2- L_1(\theta_s,\varphi)\int _0^{\pi}\sin \theta _ud\theta _u F(\theta _u) \cos (\theta _{1})\left|\cos (\theta _u) \right|.\\
\end{split}
\label{eq.3.50}
\end{equation}

Similarly, one has
\begin{equation}
\begin{split}
&\hat{E}^+_{sl  } (\theta_s,\varphi)\geq \\
&-2+ L_2(\theta_s,\varphi)\int _0^{\pi}\sin \theta _ud\theta _u F(\theta _u) \cos (\theta _{n_2})\left|\cos (\theta _u) \right|.\\
\end{split}
\label{eq.3.51}
\end{equation}

On the other hand, Eq.~(\ref{eq.3.18}) can be changed to
\begin{equation}
\begin{split}
&\hat{E}^{-}_{ll }(\varphi ')\equiv \frac{1}{N}\sum _{n=1}^NE({\bf a}^l(\frac{2n\pi}{N} + \frac{\varphi '}{2}),{\bf a}^l(\frac{2n\pi}{N} -\frac{\varphi '}{2}))\\
&\leq 2-2\cos(\frac{\varphi '}{2})\frac{1}{N}\sum _{n=1}^N\int _0^{\pi}\sin \theta _ud\theta _u \\
&\times F(\theta _u)\left|\cos (\frac{2n\pi}{N})\sin (\theta _u)\right|.\\
\end{split}
\label{eq.3.52}
\end{equation}
Using   $ \frac{1}{N}\sum _{n=1}^N \left|\cos(\frac{2n\pi}{N} + \beta)\right|\geq \frac{1}{N}\cot \left(\frac{\pi}{2N}\right)\equiv u_N$~\cite{Branciard},  one obtains
\begin{equation}
\begin{split}
&\hat{E}^{-}_{ll  }(\varphi)\leq 2-2u_N\cos (\frac{\varphi}{2})\int _0^{\pi}\sin \theta _ud\theta _u F(\theta _u) \left|\sin (\theta _u)\right|.\\
\end{split}
\label{eq.3.54}
\end{equation}

Similarly, we  have
\begin{equation}
\begin{split}
&\hat{E}^{+}_{ll  }(\varphi)\leq -2+2u_N\cos (\frac{\varphi}{2})\int _0^{\pi}\sin \theta _ud\theta _u F(\theta _u) \left|\sin (\theta _u)\right|.\\
&\hat{E}^+_{ss }(\varphi)\geq -2+2u_N\sin (\theta _s)\left|\sin (\frac{\varphi}{2})\right|\\
&\times \int _0^{\pi}\sin \theta _ud\theta _u F(\theta _u) \left|\sin (\theta _u)\right|.\\
\end{split}
\label{eq.3.55}
\end{equation}
With Eqs.~(\ref{eq.3.50}), (\ref{eq.3.51}), (\ref{eq.3.54}) and (\ref{eq.3.55}),   we  establish LI in terms of  the discrete version of average correlation functions,
\begin{equation}
\begin{split}
&\hat{E}^{-}_{sl }(\varphi _1)+\frac{\cos (\theta _{1}) L_1(\theta_s,\varphi _1)}{2u_N\cos(\frac{\varphi _2}{2})}\hat{E}^{-}_{ll }(\varphi _2)\\
&\leq 2\left(1+\frac{\cos (\theta _{1}) L_1(\theta_s,\varphi _1)}{2u_N\cos(\frac{\varphi _2}{2})}\right)-\cos (\theta _{1})L_1(\theta_s,\varphi _1),\\
&\hat{E}^+_{sl}(\varphi _1)+\frac{\cos (\theta _{2}) L_2(\theta_s,\varphi _1)}{2u_N\left|\sin(\frac{\varphi _2}{2})\right|}\hat{E}^+_{ll N}(\varphi _2)\\
&\geq -2\left(1+\frac{ \cos (\theta _{2}) L_2(\theta_s,\varphi _1)}{2u_N\left|\sin(\frac{\varphi _2}{2})\right|}\right)+\cos (\theta _2)L_2(\theta_s,\varphi _1).\\
&\hat{E}^+_{sl }(\varphi _1)+\frac{\cos (\theta _{2}) L_2(\theta_s,\varphi _1)}{2u_N\sin(\theta _s)\left|\sin(\frac{\varphi _2}{2})\right|}\hat{E}^+_{ss N}(\varphi _2)\\
&\geq -2\left(1+\frac{\cos (\theta _{2}) L_2(\theta_s,\varphi _1)}{2u_N\sin(\theta _s)\left|\sin(\frac{\varphi _2}{2})\right|}\right)+\cos (\theta _2)L_2(\theta_s,\varphi _1).\\
\end{split}
\label{eq.3.56}
\end{equation}

\end{document}